\documentclass[aps,prd,preprint,nofootinbib,a4paper,11pt,superscriptaddress]{revtex4-1}

\usepackage[T1]{fontenc}
\usepackage[centertags]{amsmath}
\usepackage{amsfonts}
\usepackage{amssymb}
\usepackage[sort&compress]{natbib}% упорядочивает ссылки
\usepackage[pdftex]{hyperref}%hypertex
\usepackage[pdftex]{graphicx}
\usepackage[russian,english]{babel}

\DeclareMathOperator{\Sp}{Sp}

\DeclareMathOperator{\tah}{th}
\DeclareMathOperator{\sgn}{sgn}

\newcommand{\lan}{\langle}
\newcommand{\ran}{\rangle}

\newcommand{\spx}{\mathbf{x}}
\newcommand{\spy}{\mathbf{y}}
\newcommand{\bnabla}{\bar{\nabla}}
\newcommand{\br}[1]{(#1)}

\newcommand{\e}{\varepsilon}
\newcommand{\vf}{\varphi}
\newcommand{\s}{\sigma}
\newcommand{\bs}{\bar{\sigma}}
\newcommand{\Si}{\Sigma}
\newcommand{\al}{\alpha}
\newcommand{\be}{\beta}
\newcommand{\ga}{\gamma}
\newcommand{\Ga}{\Gamma}
\newcommand{\de}{\delta}
\newcommand{\De}{\Delta}
\newcommand{\ka}{\varkappa}
\newcommand{\la}{\lambda}
\newcommand{\La}{\Lambda}
\newcommand{\ups}{\upsilon}

\begin{document}

%\selectlanguage{english}

\title{Propagator of a scalar field on a stationary slowly varying gravitational background}

\date{\today}

\author{P.O. Kazinski}
\email[E-mail:]{kpo@phys.tsu.ru}
\affiliation{Physics Faculty, Tomsk State University, Tomsk, 634050 Russia}
\affiliation{Department of Higher Mathematics and Mathematical Physics, Tomsk Polytechnic University, Tomsk, 634050 Russia}

\begin{abstract}

The propagator of a scalar field on a stationary slowly varying in space gravitational background is derived retaining only the second derivatives of the metric. The corresponding one-loop effective action is constructed. The propagator and the effective action turn out to depend nontrivially on the Killing vector defining the vacuum state and the Hamiltonian of a scalar field. The Hawking particle production is described in the quasiclassical approximation and the quasiclassical formula for the Hawking temperature is derived. The behaviour of the Unruh detector on a curved background is considered and the quasiclassical formula for the Unruh acceleration determining the Unruh temperature is derived. The radiation reaction problem on a curved background is discussed in view of the new approximate expression for the propagator. The correction to the mass squared of a scalar particle on a stationary gravitational background is obtained. This correction is analogous to the quantum correction to the particle mass in a strong electromagnetic field. For a vacuum solution to the Einstein equations, it is equal to minus one-fourth of the free fall acceleration squared.

\end{abstract}

%04.60.-m Quantum gravity
%04.62.+v QFT in curved spacetime

\maketitle

\section{Introduction}

In calculating the propagator of quantum fields on a fixed curved background, we encounter severe technical problems (see for review,  \cite{VasilHeatKer}) such that the exact expression can be obtained only for a very limited class of spacetimes possessing a large symmetry group. However, for many physical problems we need an approximate expression, at least, under the assumption that the gravitational field varies slowly. This is the so-called low energy approximation \cite{VasilHeatKer,Avramid}. In the present paper we shall obtain the explicit expression for the propagator of a scalar field on a stationary slowly varying in space gravitational background retaining only the second derivatives of the metric.

The essence of the problem in constructing the approximate expression for the propagator in the external gravitational field is related to the fact that the background field stands at the second derivatives entering the Klein-Gordon operator. In other words, this field enters the principal symbol of the operator \cite{BerezMSQ}. So, in the leading approximation when the quadratic part of the symbol is taken into account, the metric field should be considered as a constant matrix. Of course, such an approximation is inappropriate for sufficiently large point separations or for the problems where the next to leading corrections to the propagator depending on the derivatives of the metric are relevant. The Hawking particle production \cite{Hawk}, the Unruh effect \cite{Unruh}, and the radiation reaction problem \cite{DeWBr,Hobbs,Pois,GalSpir} on a curved background are among such physical problems. We shall derive the next to leading correction to the propagator employing appreciably the stationarity of the background metric. Of course, there are other attempts in the literature to obtain the nonperturbative expression for the propagator containing the second derivatives of the metric (see, e.g., \cite{VasilHeatKer,Avramid,BekPark,Parker,Page,BarVilkcov,WiMuZe,HuOCon}). However, none of these approaches can reproduce the Hawking particle production for four dimensional spacetimes with sufficiently general metrics. The reasons for this are as follows:
\renewcommand{\theenumi}{\roman{enumi}}
\begin{enumerate}
  \item The most of these papers treat the problem in a Riemann spacetime instead of a pseudo-Riemann one and make then an ``analytical continuation'' back to the spacetime with Lorentzian signature metric. So, it is even unclear what the quantum state determining the propagator stands behind such a procedure. Usually it is assumed that it is the Hartle-Hawking vacuum, but when the spacetime has no horizons the definition of such a vacuum state is obscure;
  \item For static metrics there is a more well-founded approach \cite{Page} based on the conformal transformations technique. Using the conformal transformation, one can pass to the optic (ultrastatic) metric where the time variable completely decouples. Then the methods \cite{BekPark} elaborated for the operators of Laplacian type are applied in the spatial sector leading to the approximate expression for the propagator. Eventually, the inverse conformal transform is performed. We shall discuss such an approach in some details in this paper and show that, for strong gravitational fields, it is very sensitive to the higher order corrections to the leading asymptotics derived in \cite{Page} and so cannot reproduce the Hawking particle production and other exponentially suppressed contributions to the propagator.
\end{enumerate}
On the other hand, it is clear even by dimensional reasons that the properly constructed propagator, where only the second derivatives of the metric are taken into account, must reproduce the Hawking radiation. Bearing in mind the analogy of the Hawking particle production with the Unruh effect, one may expect that the Hawking temperature should be determined by the ``acceleration squared'' and, consequently, must be expressed in terms of the second order derivatives of the metric field. In the present paper, we shall obtain the approximate expression for the propagator associated with the standard vacuum state for stationary backgrounds (see \cite{DeWGAQFT}, Chap. 17) by a brute force calculation. The main idea is quite plain: We make a Fourier transform of the Klein-Gordon equation with respect to the time variable and obtain the operator of Laplacian type (rather than of the hyperbolic one) depending parametrically on the frequency. After that we apply the developed methods to this Fourier transformed operator and derive thereby the approximate expression for the Fourier transformed propagator. At the end, the inverse Fourier transform is performed. The result will reproduce the Hawking particle production and also other exponentially suppressed contributions. Besides, the propagator will explicitly depend on the Killing vector field determining the vacuum state.

It was shown in \cite{gmse} (see also \cite{BrOtPa,FrZel}) that for stationary backgrounds the Killing vector $\xi^\mu$ appears in the effective action in a nontrivial way, that is the expressions depending on it cannot be rewritten in terms of the local expressions involving the metric and the curvature alone. This fact can be used as another one method to measure experimentally the variation of the Killing vector (up to multiplication by a constant) from point to point. The present paper continues the research in this direction. We are about to investigate in detail how the Killing vector field influences (determines) the quantum dynamics of the matter fields. The first object for such an investigation is, of course, the two-point Green function. Making use of the explicit expression for the propagator, we shall derive the analog of the Heisenberg-Euler effective action \cite{Schwing,HeisEul} which, as we shall see, depends nontrivially on the vector field $\xi^\mu$. The imaginary part of the effective action, which is responsible for the Hawking radiation, turns out to depend nontrivially on the Killing vector and cannot be removed by the real counterterms.  The quasiclassical formulas for the Hawking particle production \cite{Hawk} and for the Unruh temperature \cite{Unruh} of a detector moving in a stationary gravitational field will be also obtained. Besides, we shall find the quantum correction to the mass of a particle. This correction is not due to the Higgs mechanism, but a direct one, and similar to the correction to the mass of a charged particle in a strong electromagnetic field \cite{Ritus2p}. On the vacuum solutions to the Einstein equations ($R_{\mu\nu}=0$), the correction to the mass squared is found to be equal to minus one-fourth of the free fall acceleration squared. Such a correction to mass is also expectable if one recalls that the gravitational field produces particles at the Hawking temperature. The analogous corrections to the mass squared are well known in quantum field theory at finite temperature \cite{DolJack,GrPeYa,Kapusta}.

Further, in the main body of the paper, we shall also discuss the radiation reaction effect on a curved background as a possible method to measure the direction of the Killing vector \cite{DeWBr}. The fact that the Killing vector field of a stationary metric can be measured experimentally is more or less known. It is related with our ability to determine experimentally that the particle (detector) is at rest in a stationary gravitational field. Undoubtedly, in the presence of gravity the resting state of a particle is not a free fall along geodesic. Hence, there must be a certain structure on a spacetime that distinguishes the worldlines of resting particles. If we know such trajectories, we can send the light signals between the detectors moving along these worldlines and, making a comparison the red(blue)shifts (the standard redshift formula is assumed, see, e.g., \cite{Wald}, Chap. 6; \cite{LandLifshCTF}, Sec. 88), restore the covariant square of the Killing vector at any point of space up to an overall constant factor. The Killing vector is then obtained as $\xi^\mu=|\xi^2|^{1/2}u^\mu$, where $u^\mu$ is the four-velocity of a detector. One of the possible means how to determine that the particle is at rest can be as follows. Consider a charged particle in a stationary spacetime. If the Killing vector defining the stationarity of the metric is tangent to the worldline of this particle then the particle does not radiate (the radiation is measured at the null infinity where the spacetime is supposed to be flat). The absence of radiation in this case follows from the fact that the translation in time is the symmetry of the system (particle)+(gravitational field). On the other worldlines of a general form and, in particular, for the geodesic motion, the charged particle will radiate as can be shown perturbatively, at least, in the weak field limit.

Of course, one may contrive less sophisticated methods to measure the various contractions of the Killing vector \cite{Okun,DeWdet,Anandan,Will}. Especially promising experiments are those that involve macroscopic quantum effects (see, e.g., \cite{KasChu}). The explicit dependence on the Killing vector is inherent to all quantum phenomena in contrast to the ideal classical mechanical systems of point particles obeying the geodesic equations of motion. The latter are determined solely by the metric. We address just the above mentioned problems (the Hawking particle production, the Unruh effect, and the radiation reaction problem) as their solutions follow immediately from the form of the propagator. It should be noted that the latter two problems are widely discussed in the literature arriving at the contradictory conclusions (see, e.g., \cite{GinzLogrr,NFKMFU}). One of the aims of the present paper is to provide a more solid mathematical background for these considerations on a curved background.

One can provide more rough definitions of the resting system of coordinates associated with the Killing vector and the geodetic (free fall) one such that their distinction becomes evident. In a vicinity of a spacetime point, we define the resting frame as the system of coordinates where the uncharged particles fall with the acceleration which does not depend on time in a fixed point of space. In the geodetic system of coordinates, the particles fall without acceleration at the origin. These systems of coordinates coincide in the absence of the gravitational field. The reader may ask where the quantum apparatuses in these definitions which distinguish the resting and geodetic frames. The answer is these devices are the clocks defining the time in these systems of coordinates. The clocks are supposed to be quantum, for example, the atomic clock. For comparison one can consider the above definitions with the purely classical clock -- the pendulum. With respect to the time counted by the atomic clock, the pendulum ticks with the frequency proportional $a^{1/2}$, where $a$ is the free fall acceleration. Therefore, the acceleration of a body in a free fall measured with respect to the pendulum clock is the same both for sufficiently large gravitational field and for very small one and is independent of time. The limit of very small gravitational fields corresponds to the geodetic system of coordinates. We see that near the origin the two systems of coordinates described above are classically indistinguishable in a complete agreement with the equivalence principle for purely classical systems.

The paper is organized as follows. In Sec. \ref{Asympt_Expans}, the derivative expansion of the Feynman propagator is considered near its diagonal up to the second derivatives of the stationary metric field. The stationarity of the metric is essentially employed. After the Fourier transform of the Klein-Gordon operator with respect to the time variable the generalized Schwinger-DeWitt technique \cite{BarvVilk} adapted to the $(3+1)$-decomposition is applied. We shall see that the simple derivative expansion has a large degree of ambiguity and cannot be used to restore the propagator unambiguously. In Sec. \ref{Resum_Expansion}, the derivative expansion is resummed making use of the Schwinger approach \cite{Schwing}, and the low energy expansion \cite{Avramid,VasilHeatKer} of the propagator is constructed thereby. There we shall derive the explicit expression for the propagator (or, what is equivalent, for the positive-frequency function) as the double integral over the frequencies of the modes and an auxiliary variable -- the Fock proper-time \cite{Fock}. In fact, the results of Sec. \ref{Asympt_Expans} are only needed to crosscheck the nonperturbative expression for the propagator obtained in Sec. \ref{Resum_Expansion} and to derive the only one higher derivative term which is not reproduced in the leading order of the approximate procedure used in Sec. \ref{Resum_Expansion}. Section \ref{Phys_Implic} is devoted to a further simplification of the expression for the positive-frequency function under the assumption that the wavelengths of modes, the propagation of which the constructed propagator should describe, are much less than characteristic scale of variations of the gravitational field. We shall obtain the spectral representation of the propagator and its representation in terms of the proper-time. The expression for the one-loop correction to the effective action will be also derived. In conclusion, we shall discuss some implications of the results of the present paper and the possible generalization to non-stationary spacetimes. In Appendix \ref{4_Kill}, several formulas concerning the calculus on the spacetime with the Killing vector are presented. There, in particular, the expressions for the $4$-vectors of the gravitational and inertial forces constructed with the help of the Killing vector field $\xi^\mu$ will be given. Some useful expansions needed in the course of derivation of the approximate expression for the propagator are collected in Appendix \ref{Some_Expan}. In Appendix \ref{Conf_Trans}, we shall discuss the conformal transformation technique and estimate the higher derivative corrections to the propagator for different conformal factors.

We shall use the conventions adopted in \cite{DeWGAQFT,BarvVilk}
\begin{equation}
    R^\al_{\ \be\mu\nu}=\partial_{[\mu}\Ga^\al_{\nu]\beta}+\Ga^\al_{[\mu\ga}\Ga^\ga_{\nu]\beta},\qquad R_{\mu\nu}=R^\al_{\ \mu\al\nu},\qquad R=R^\mu_\mu,
\end{equation}
for the curvatures and other structures appearing in the heat kernel expansion. The square and round brackets at a pair of indices denote antisymmetrization and symmetrization without $1/2$, respectively. The Greek indices are raised and lowered by the metric $g_{\mu\nu}$ which has the signature $+2$. Also we assume that the metric possesses the timelike Killing vector $\xi^\mu$
\begin{equation}
    \mathcal{L}_\xi g_{\mu\nu}=0,\qquad\xi^2=g_{\mu\nu}\xi^\mu\xi^\nu<0,
\end{equation}
that allows us to make the decomposition \cite{LandLifshCTF}, Sec. 84,
\begin{equation}\label{metric_decomp}
    ds^2=g_{\mu\nu}dx^\mu dx^\nu=:\xi^2(g_\mu dx^\mu)^2+\bar{g}_{\mu\nu}dx^\mu dx^\nu,
\end{equation}
where $g_\mu=\xi_\mu/\xi^2$ is a one-form dual to the Killing vector (the Tolman temperature one-form). In the system of coordinates, where $\xi^\mu=(1,0,0,0)$, we have the relations
\begin{equation}
\begin{gathered}
    \bar{g}_{ik}g^{kj}=\de_i^j,\qquad \xi^2\det\bar{g}_{ij}=g,\qquad g^{00}-\bar{g}_{ij}g^{0i}g^{0j}=(g_{00})^{-1},\\
    \bar{g}_{\mu\nu}=\begin{bmatrix}
                       0 & 0 \\
                       0 & g_{ij}-\frac{g_{i0}g_{j0}}{g_{00}} \\
                     \end{bmatrix},\qquad
    \bar{g}^{\mu\nu}=g^{\mu\nu}-\xi^2g^\mu g^\nu=\begin{bmatrix}
                       g^{00}-(g_{00})^{-1} & g^{0j} \\
                       g^{i0} & g^{ij} \\
                     \end{bmatrix}.
\end{gathered}
\end{equation}
The Latin indices corresponding to the space are raised and lowered by the positive-definite metric $\bar{g}_{ij}$. The curvatures associated with this metric will be distinguished by the overbars, e.g., $\bar{R}$. Note that we consider a general stationary spacetime, i.e., the Tolman temperature one-form is supposed to be non-integrable (\cite{Wald}, App. C; \cite{LandLifshCTF}, Sec. 88):
\begin{equation}
    f_{\mu\nu}:=\partial_{[\mu}g_{\nu]}\neq0,
\end{equation}
in general. The system of units is chosen such that $c=\hbar=1$.

%\newpage
\section{Asymptotic expansion of the propagator near the diagonal}\label{Asympt_Expans}

Our aim in this section is to obtain the derivative expansion of the Feynman propagator near its diagonal up to the second derivatives of the stationary metric field $g_{\mu\nu}$. We shall take exactly into account the symmetry of the background metric with respect to translations in time and employ the so-called generalized Schwinger-DeWitt technique \cite{BarvVilk} adapted to the $(3+1)$-decomposition. Hereinafter, we shall work in the reference frame where the Killing vector is straighten ($\xi^\mu=(1,0,0,0)$). Besides, all the formulas presented in this section are valid for the arbitrary spacetime dimension $(d+1)$.

Let
\begin{multline}\label{KG}
    H(x,y)=(\nabla^2_x-m^2)\frac{\de(x-y)}{|g|^{1/4}(x)|g|^{1/4}(y)}=\\
    =|g|^{-1/4}(x)\biggl[|g|^{-1/4}(x)\partial_\mu\sqrt{|g|}g^{\mu\nu}\partial_\nu|g|^{-1/4}(x)-m^2\biggr]\frac{\de(x-y)}{|g|^{1/4}(y)}
\end{multline}
be the bi-scalar kernel of the Klein-Gordon operator. The mass squared $m^2$ is assumed to be constant, i.e., we consider a free scalar field on a curved background. The generalization to the case where the mass is generated by the Higgs mechanism, and consequently may depend on the spacetime point, is straightforward. Supposing the metric does not depend on time $t$, we can rewrite \eqref{KG} in terms of its Fourier transform
\begin{equation}\label{Hamiltonian}
\begin{split}
    H(x,y)&=:|\xi^2|^{-1/4}(\spx)\int\frac{d\omega}{2\pi}e^{-i\omega t}H(\omega;\spx,\spy)|\xi^2|^{-1/4}(\spy),\\
    H(\omega;\spx,\spy)&=\biggl[-\frac{\omega^2}{\xi^2}+(\bar{\nabla}_i+i\omega g_i)^2-m^2-\frac12\bnabla^ih_i-\frac14h_ih^i\biggr]\frac{\de(\spx-\spy)}{\bar{g}^{1/4}(\spx)\bar{g}^{1/4}(\spy)},
\end{split}
\end{equation}
where $x=(t,\spx)$ and $y=(0,\spy)$, the Levi-Civita connection $\bnabla_i$ is constructed in terms of the metric $\bar{g}_{ij}$, and we have introduced the notation $h_\mu:=\partial_\mu\ln\sqrt{|\xi^2|}$. Notice that the one-form $g_i$ plays the role of a $U(1)$ gauge field with the charge $\omega$ (the energy). The kernel $H(\omega;\spx,\spy)$ is a bi-scalar with respect to the spatial diffeomorphisms. Then the positive-frequency function defined with respect to the standard vacuum state for stationary backgrounds (in black hole physics it is known as the Boulware vacuum \cite{Boulware}) is written as
\begin{equation}\label{pos_freq_func}
\begin{gathered}
    D^{(+)}(x,y):=-i\lan\phi(x)\phi(y)\ran=-i\int_0^\infty d\omega e^{-i\omega t_-}|\xi^2|^{-1/4}(\spx)\lan \spx|\de(H(\omega))|\spy\ran|\xi^2|^{-1/4}(\spy),\\
    \lan\spx|\de(H(\omega))|\spy\ran=\int\frac{ds}{2\pi}\lan\spx|e^{isH(\omega)}|\spy\ran=:\int \frac{ds}{2\pi} G(\omega,s;\spx,\spy),
\end{gathered}
\end{equation}
where $t_-:=t-i0$ and the integration contour in the $s$-plane passes a little bit lower than the real axis. The Feynman propagator is expressed in terms of the positive-frequency function: at $t>0$ the propagator coincides with it, while at $t<0$ we should make the replacement,
\begin{equation}
    t_-\rightarrow-t+i0,\qquad\spx\leftrightarrow\spy,
\end{equation}
in $D^{(+)}(x,y)$. So, it is sufficient to calculate the positive-frequency function only.

To this end, we can apply the generalized Schwinger-DeWitt technique to the heat kernel $G(\omega,s;\spx,\spy)$ assuming that the points $\spx$ and $\spy$ can be connected by a unique geodesic of the metric $\bar{g}_{ij}$. Making the derivative expansion, it is convenient to use the covariant $qp$-symbol of the heat kernel \cite{Kuzen,BerezMSQ,BarvVilk,DeWGAQFT}
\begin{multline}\label{HK_qp}
    G=\exp\biggl\{is(-\frac{\omega^2}{\xi^2}+D^2_i-m^2-\frac12\bnabla^ih_i-\frac14h_ih^i)\biggr\}\int\frac{d^dk'}{(2\pi)^d}\sqrt{\bar{g}'}e^{ik^{i'}\bs_{i'}}a_0(\spx,\spy)=\\
    =\int\frac{d^dk'}{(2\pi)^d}\sqrt{\bar{g}'}e^{ik^{i'}\bs_{i'}-is(\frac{\omega^2}{\xi^2}+k^{i'}\bs_{i'i}\bs^i_{\ j'}k^{j'}+m^2+\frac12\bnabla^ih_i+\frac14h_ih^i)}a_0(\spx,\spy)\times\\
    \times a_0(\spy,\spx) e^{is(\frac{\omega^2}{\xi^2}+k^{i'}\bs_{i'i}\bs^i_{\ j'}k^{j'}+\frac12\bnabla^ih_i+\frac14h_ih^i)} e^{is(-\frac{\omega^2}{\xi^2}+(D_i+ik^{i'}\bs_{i'i})^2-\frac12\bnabla^ih_i-\frac14h_ih^i)}a_0(\spx,\spy),
\end{multline}
where the primes denote the quantities and derivatives referring to the point $\spy$, the indices without prime refer to the point $\spx$, for example,
\begin{equation}
    \bs_{i'i}\equiv\bnabla_{i'}\bnabla_i\bs(\spy,\spx)=\frac{\partial^2\bs(\spy,\spx)}{\partial y^{i'}\partial x^i},
\end{equation}
where $\bs(\spy,\spx)=\bs(\spx,\spy)$ is the world function associated with the metric $\bar{g}_{ij}$. Also $a_0(\spx,\spy)$ is the geodetic parallel displacement operator with respect to the connection $D_i=\bnabla_i+i\omega g_i$,
\begin{equation}
    a_0(\spx,\spy)=e^{-i\omega\int_{\spy}^{\spx} dx^ig_i}.
\end{equation}
The integral is taken along the geodesic of the metric $\bar{g}_{ij}$ connecting the points $\spx$ and $\spy$.

The derivative expansion is obtained expanding the last line in \eqref{HK_qp} in a covariant Taylor series near the point $\spx$. At first, all the exponents in this line are expanded in the series. After that the covariant derivatives $D_i^2$ are carried through the expression to the right (using the Leibnitz rule) to act on $a_0$, whereas the derivatives of $a_0$ are expanded in the covariant Taylor series by the formulas (4.26)-(4.29) in \cite{BarvVilk}. In the notation of \cite{BarvVilk},
\begin{equation}
    \hat{\mathcal{R}}_{\mu\nu}\rightarrow i\omega f_{ij},
\end{equation}
where $f_{ij}:=\partial_{[i}g_{j]}$. The first exponent in the last line of Eq. \eqref{HK_qp} provides considerable cancelations in the expansion -- all the terms, which do not contain the derivatives of the expression standing in the exponent, disappear. The exponent in the second line effectively sums such terms (see for the proof \cite{Parker,KalKaz}). As a result, we obtain the series containing the nonnegative integer powers of $k^{i'}$. Then the Gaussian integral over $k^{i'}$ is performed. In fact, one can replace $k^{i'}$ by $-i\partial/\partial\bs_{i'}$ acting on the result of the Gaussian integration of the expression in the second line in \eqref{HK_qp}:
\begin{equation}
    \frac{\bar{\De}^{-1}(\spx,\spy)}{(4\pi is)^{d/2}}e^{\frac{i}{4s}\bs^{i'}\bs^{-1}_{i'i}\bar{g}^{ij}\bs^{-1}_{jj'}\bs^{j'}-is(m^2+\frac{\omega^2}{\xi^2}+\frac12\bnabla^ih_i+\frac14h_ih^i)}a_0(\spx,\spy).
\end{equation}
After rather lengthy but straightforward calculations one derives
\begin{multline}\label{HK_qp_expand}
    G=\frac{\bar{\De}^{-1}(\spx,\spy)}{(4\pi is)^{d/2}}e^{\frac{i}{2s}\bs-is(m^2+\frac{\omega^2}{\xi^2}+\frac12\bnabla^ih_i+\frac14h_ih^i)}a_0(\spx,\spy)\biggl\{1+\frac14 \bar{R}_{ij}\bs^i\bs^j+\frac{is}6 \bar{R}-\\
    -\frac{s^2}2\Bigl[\frac{\omega^2}{\xi^2}\Bigl(\frac23\bnabla^ih_i-\frac43 h_ih^i\Bigr)-\frac{\omega}{3s}\bnabla^jf_{ji}\bs^i-\frac{\omega^2}2f_{ij}f^{ij}+\frac{2i}{s}\frac{\omega^2}{\xi^2}h_i\bs^i\Bigr]-\\
    -\frac{is^3}{3}\Bigl[\frac{\omega^4}{\xi^4}h_ih^i+\frac{\omega^3}{s\xi^2}\bs^if_{ij}h^j+\frac{2\omega^2}{s^2\xi^2}\Bigl( (h_i\bs^i)^2-\frac12\bnabla_ih_j\bs^i\bs^j \Bigr)+\frac{\omega^2}{2is}f_{ij}f^{ij}-\frac{\omega^2}{4s^2}\bs^if_{ik}f^k_{\ j}\bs^j\Bigr]-\frac{s^2\omega^4}{2\xi^4}(h_i\bs^i)^2\biggr\}
\end{multline}
up to the second order derivatives of the background fields, all the fields being taken at the point $\spx$. Notice that in this approach the order of the term in derivatives is counted as the total number of derivatives entering it. For example, the term
\begin{equation}\label{h4}
    (h_ih^i)^2
\end{equation}
is of the fourth order in derivatives.

The heat kernel $G(\omega,s;\spx,\spy)$ is a unitary operator with respect to the standard measure $\bar{g}^{1/2}$ for real $s$, and so it is Hermitian with respect to this measure when $s$ is purely imaginary. It is desirable to preserve this symmetry in the approximate expression for the heat kernel. The expression \eqref{HK_qp_expand} is not Hermitian at $s=i\tau$ since the points $\spx$ and $\spy$ enter asymmetrically to it. Therefore, we expand \eqref{HK_qp_expand} in the vicinity of the point $p(\spx,\spy)$ lying at the middle of the geodesic connecting the points $\spx$ and $\spy$ (the midpoint prescription) and retain only the terms which are no more than of the second order in derivatives of the fields. An inspection of the expansion \eqref{HK_qp_expand} shows up that we only need to expand $\xi^{-2}$ in the exponent and the last term in the second line, while the fields entering the other terms are simply taken at the midpoint. This gives
\begin{multline}\label{HK_qp_expand2}
    G=\frac{a_0(\spx,\spy)}{(4\pi is)^{d/2}}e^{\frac{i}{2s}\bs-is(m^2+\frac{\omega^2}{\xi^2}+\frac12\bnabla^ih_i+\frac14h_ih^i)}\biggl\{1+\frac1{12} \bar{R}_{ij}\bs^i\bs^j+\\
    +\frac{is}6 \bar{R}+\frac{is}{12}\frac{\omega^2}{\xi^2}(\bnabla_ih_j-2h_ih_j)\bs^i\bs^j
    -\frac{s^2}6\Bigl[\frac{\omega^2}{\xi^2}\Bigl(2\bnabla^ih_i-4 h_ih^i\Bigr)-\frac{\omega}{s}\bnabla^jf_{ji}\bs^i-\frac{\omega^2}2f_{ij}f^{ij}\Bigr]-\\
    -\frac{is^3}{3}\Bigl[\frac{\omega^4}{\xi^4}h_ih^i+\frac{\omega^3}{s\xi^2}\bs^if_{ij}h^j-\frac{\omega^2}{4s^2}\bs^if_{ik}f^k_{\ j}\bs^j\Bigr]\biggr\},
\end{multline}
where we also expand the bi-scalar van Vleck determinant $\bar{\De}(\spx,\spy)$ (see Appendix \ref{Some_Expan}).

Now we should substitute the expansion obtained to the integral \eqref{pos_freq_func} and integrate it over $\omega$ and $s$. However, at this point we have a large degree of ambiguity how to represent the substituting expression, and this ambiguity can substantially change the resulting positive-frequency function. For example, we may expand the exponent in \eqref{HK_qp_expand2} and neglect the higher order terms or, conversely, collect some terms in the preexponential factor to the exponent similarly to what we did with the term $\omega^2/\xi^2$. The naive derivative expansion, which we carried out in this section, does not tell us what the right form of the integrand is. We shall address this problem in the next section, while here we establish the relation of the expansion \eqref{HK_qp_expand2} with the standard one \cite{DeWGAQFT,BarvVilk} written in terms of the geodetic interval of the spacetime rather than $\bs$. In such a way, we shall make a nontrivial check of the correctness of the expression \eqref{HK_qp_expand2}.

In order to obtain the standard asymptotic expansion of the Feynman propagator, it is convenient to use the Schwinger representation for it
\begin{equation}\label{Schwing_repres}
\begin{gathered}
    D(x,y):=-i\lan T\{\phi(x)\phi(y)\}\ran=\int_{-\infty}^\infty \frac{d\omega}{2\pi} e^{-i\omega t}|\xi^2|^{-1/4}(\spx)\lan \spx|(H(\omega)+i0)^{-1}|\spy\ran|\xi^2|^{-1/4}(\spy),\\
    \lan\spx|(H(\omega)+i0)^{-1}|\spy\ran=\int_0^{\infty} ids\lan\spx|e^{is(H(\omega)+i0)}|\spy\ran=\int_0^{\infty} ids G(\omega,s;\spx,\spy)|_{m^2\rightarrow m^2-i0},
\end{gathered}
\end{equation}
instead of using Eq. \eqref{pos_freq_func}. Then we can formally integrate over $\omega$ the expansion \eqref{HK_qp_expand2} as the integral of a Gaussian type. This leads to
\begin{multline}\label{Schwing_repres_1}
    \int\frac{id\omega}{2\pi}e^{-i\omega t}|\xi^2|^{-1/4}G|\xi'^2|^{-1/4}\approx\frac{ie^{\frac{i}{4s}(\xi^2T^2+2\bs)-ism^2}}{(4\pi is)^{(d+1)/2}}\biggl\{\frac{i}{48s}\Bigl[\xi^2T^2(\bnabla_ih_j-2h_ih_j+\xi^2f_{ik}f^k_{\ j})\bs^i\bs^j+\\
    +2\xi^4T^3\bs^if_{ij}h^j-\xi^4T^4h^2 \Bigr]+1+\frac1{12}\Bigl[\Bigl(\bar{R}_{ij}-h_ih_j-\bnabla_ih_j+\frac{\xi^2}2f_{ik}f^k_{\ j}\Bigr)\bs^i\bs^j+\xi^2T^2(2h^2-\bnabla^ih_i)-\\
    -\xi^2T\bnabla^jf_{ji}\bs^i+\frac14\xi^4T^2f^2
    -3\xi^2T^2h^2+3\xi^2T\bs^if_{ij}h^j \Bigr]+\frac{is}6\Bigl[\bar{R}-2h^2-2\bnabla^ih_i-\frac{\xi^2}4f^2\Bigr]\bigg\},
\end{multline}
where $T:=t+\int_{\spy}^{\spx} dx^ig_i$ is a scalar with respect to the general coordinate transformations \cite{LandLifshCTF}, Sec. 88. To obtain \eqref{Schwing_repres_1} we have used the expansions from Appendix \ref{Some_Expan}. Bearing in mind that $\bs_\mu=0$ at $\mu=0$ in the adapted system of coordinates (i.e. $\xi^\mu\bs_\mu=0$) and applying the formulas given in Appendix \ref{4_Kill}, we can write to the same accuracy
\begin{multline}\label{HK_stand}
    \int\frac{id\omega}{2\pi}e^{-i\omega t}|\xi^2|^{-1/4}G|\xi'^2|^{-1/4}\approx\frac{ie^{\frac{i}{4s}(\xi^2T^2+2\bs)-ism^2}}{(4\pi is)^{(d+1)/2}}\Big\{-\frac{i\xi^2T^2}{48s}\bigl[\xi^2(Th_\mu+f_{\mu\nu}\bs^\nu)^2-\\
    -(\nabla_\mu h_\nu-2h_\mu h_\nu)\bs^\mu\bs^\nu \bigr]
    +1+\frac1{12} R_{\mu\nu}(T\xi^\mu+\bs^\mu)(T\xi^\nu+\bs^\nu)+\frac{is}6R\Big\}\approx\\
    \approx \exp\Big[\frac{i}{4s}\big\{\xi^2T^2+2\bs-\frac{\xi^2T^2}{12}[\xi^2(Th_\mu+f_{\mu\nu}\bs^\nu)^2-(\nabla_\mu h_\nu-2h_\mu h_\nu)\bs^\mu\bs^\nu ]\big\}-ism^2\Big]\times\\
    \times\frac{i}{(4\pi is)^{(d+1)/2}}\Big\{1+\frac1{12} R_{\mu\nu}(T\xi^\mu+\bs^\mu)(T\xi^\nu+\bs^\nu)+\frac{is}6R\Big\}.
\end{multline}
Up to the terms of the second order in derivatives, the quantity standing in the curly brackets in the exponent is the geodetic interval squared, and
\begin{equation}
    T\xi_\mu+\bs_\mu\approx\s_\mu
\end{equation}
to the leading order in derivatives.

Indeed, it is sufficient to consider
\begin{equation}\label{quasiinterval}
    X^2(x,y):=\xi^2(p)T^2+2\bs-\frac{\xi^2T^2}{12}[\xi^2(Th_\mu+f_{\mu\nu}\bs^\nu)^2-(\nabla_\mu h_\nu-2h_\mu h_\nu)\bs^\mu\bs^\nu ]
\end{equation}
on the arbitrary worldline $x^\mu(\tau)$, where $\tau$ is the natural parameter
\begin{equation}\label{natural_param}
    g_{\mu\nu}(x(\tau))\dot{x}^\mu(\tau)\dot{x}^\nu(\tau)=-1.
\end{equation}
For definiteness, we can take $x^\mu(0)=y=0$ with zero being the origin of the Riemann normal coordinates of the metric $\bar{g}_{ij}$ (see for details \cite{Petrov}). To fix the frame in the spacetime uniquely, it is convenient to use the Fock gauge \cite{Zelman,LandLifshCTF,Kuzen}
\begin{equation}\label{Fock_gauge}
    g_i(x)x^i=0\;\Leftrightarrow\;g_i(x)=\sum_{n=1}^\infty\frac{n}{(n+1)!}x^{j_1}\cdots x^{j_n}\partial_{j_1}\ldots\partial_{j_{n-1}}f_{j_ni}=\frac12\bs^jf_{ji}+\frac13\bs^{j_1}\bs^{j_2}\bnabla_{j_1}f_{j_2i}+\ldots,
\end{equation}
where $x^i$ are the coordinates of the Riemann frame. This gauge is equivalent to
\begin{equation}
    \int_0^{\spx} dx^ig_i=0,
\end{equation}
for any $x^i$ of the Riemann normal coordinates, where the integral is taken along a straight line. In this system of coordinates $T=t$. Note that the higher terms of the expansion \eqref{Fock_gauge} depend on the curvature \cite{Petrov}, but they will be irrelevant in our case as they are of the higher order in derivatives. Then expanding the expression \eqref{quasiinterval} in $\tau$ and making use of the relations \eqref{natural_param}, \eqref{Fock_gauge}, one arrives at (see Appendix \ref{Some_Expan})
\begin{equation}\label{quasiint_exp}
    X^2=-\tau^2-\frac1{12}(\nabla_uu)_{\tau=0}^2\tau^4+\ldots,\qquad u^\mu:=\dot{x}^\mu(\tau),
\end{equation}
up to the terms of the second order in derivatives of the metric $g_{\mu\nu}$. The Killing vector disappears from the expression. In a free fall, $X^2=-\tau^2$ and, consequently, with the accuracy we work, $X^2$ coincides with the geodetic interval squared. Thus we see that the expansion \eqref{HK_stand} agrees with the standard asymptotic expansion of the heat kernel near the diagonal \cite{DeWGAQFT,BarvVilk}. So, the expansion of the propagator following from \eqref{Schwing_repres} also coincides with the standard one up to the terms of the second order in derivatives.

%\newpage
\section{Resummation of the expansion}\label{Resum_Expansion}

In the previous section we saw that the resulting expression for the positive-frequency function $D^{(+)}(x,y)$ depends severely on rearrangements of the derivative expansion. The very fact that we was able to integrate expansion \eqref{HK_qp_expand2} over the frequency $\omega$ is a consequence of the resummation of the derivative expansion, which we have done raising the potential term $\omega^2/\xi^2$ to the exponent. Therefore we need a more reliable procedure to obtain an approximate but adequate expression for the propagator. This procedure should sum an infinite number of terms of the derivative expansion, say, all the terms containing the fields at the given point, their first, second derivatives and not the third derivatives and higher. This is the so-called low energy expansion of the heat kernel \cite{Avramid,VasilHeatKer}. For example, in this case the term of the form \eqref{h4} must not be neglected. In this section, we obtain the leading term in the low energy expansion of the heat kernel slightly generalizing the standard procedure \cite{Schwing} to the curved spacetime. Now we put $d=3$.

To begin with, we consider a simple model in a flat Euclidean space with the action
\begin{equation}\label{model_simpl}
    S[x(\tau)]=\int_0^s d\tau\Bigl( \frac12\dot{x}^2-A_i(x)\dot{x}^i+\frac12E_{ij}x^ix^j\Bigr),
\end{equation}
where $A_i=x^jf_{ji}/2$, the field strength matrix $f_{ij}$ is constant and skewsymmetric, and $E_{ij}$ is a constant symmetric matrix. This is a general quadratic model and later on we shall see how to use it to construct a reliable positive-frequency function of the scalar field on a curved background. Further, we assume
\begin{equation}\label{E_f}
    [E,f]=0\;\Rightarrow\; f=-iH\ups^1_{[i}\bar{\ups}^1_{j]}=:-iHa_{ij},\quad E=\la_1\ups^1_{(i}\bar{\ups}^1_{j)}+\la_2\ups^2_i\ups^2_j=:\la_1s_{ij}+\la_2\ups^2_i\ups^2_j.
\end{equation}
The vectors $\ups^1_i$, $\bar{\ups}^1_i$, and $\ups^2_i$ are orthonormal with respect to the standard Hermitian scalar product, the overbar denotes complex conjugation, the vector $\ups^2_i$ having real components. In this case, the equations of motion for this model can be readily integrated
\begin{equation}\label{eqm_sol}
    x=\bigl[(\zeta_1e^{-i\omega^+\tau}+\zeta_2e^{-i\omega^-\tau})\ups^1+c.c.\bigr]+(c_1e^{\sqrt{\la_2}\tau}+c_2e^{-\sqrt{\la_2}\tau})\ups^2,\qquad
    \omega^\pm:=\frac12(H\pm\sqrt{H^2-4\la_1}),
\end{equation}
where $\zeta_1$ and $\zeta_2$ are the complex numbers, $c_1$ and $c_2$ are real, and we assume for definiteness that $\la_2>0$ and $\la_1<0$. This is indeed the case for the gravitational field (see below). The constants of integration are uniquely determined by the boundary conditions $x(0)=y$ and $x(s)=x$. Then the Hamilton-Jacobi action takes the form:
\begin{multline}\label{Ham_Jac_action_pre}
    S=\frac12(x_i(s)\dot{x}^i(s)-x_i(0)\dot{x}^i(0))
    =\frac14(y^is_{ij}y^j+x^is_{ij}x^j)\sqrt{H^2-4\la_1}\ctg\frac{s}2\sqrt{H^2-4\la_1}-\\
    -\frac{\sqrt{H^2-4\la_1}}{2\sin\frac{s}2\sqrt{H^2-4\la_1}}(x^is_{ij}y^j\cos\frac{s}2H+ix^ia_{ij}y^j\sin\frac{s}2H )+\\
    +\frac{\sqrt{\la_2}}2\cth s\sqrt{\la_2}[(\ups^2_iy^i)^2+(\ups^2_ix^i)^2]-\frac{\sqrt{\la_2}}{\sh s\sqrt{\la_2}}x^i\ups_i^2\ups_j^2y^j%=\\
    =\frac14y\sqrt{-f^2-4E}\ctg\frac{s}2\sqrt{-f^2-4E}y+\\
    +\frac14x\sqrt{-f^2-4E}\ctg\frac{s}2\sqrt{-f^2-4E}x-y\frac{\sqrt{-f^2-4E}e^{\frac{s}2f}}{2\sin\frac{s}2\sqrt{-f^2-4E}}x.
\end{multline}
Shifting the variables $x^i$ in the initial action \eqref{model_simpl} by a constant vector, it is easy to generalize the above result to the case where the Lagrangian contains the additional terms $E_0$ and $b_ix^i$. And so, for the system with the Hamilton function
\begin{equation}\label{Ham_func}
    H(p,x)=(p_i+A_i)^2-E_0-b_ix^i-\frac{1}{2}E_{ij}x^ix^j,
\end{equation}
we have the Hamilton-Jacobi action
\begin{multline}\label{Ham_Jac_action}
    S=\frac14(x-y)\ka\ctg s\ka(x-y)+\tilde{y}\frac{\ka}{2}(\ctg s\ka-\frac{e^{sf}}{\sin s\ka})\tilde{x}-\frac{s}2bE^{-1}b+\frac12bE^{-1}f(x-y)+sE_0,
\end{multline}
where $\ka:=\sqrt{-f^2-2E}$, and the shifted variables $\tilde{x}_i=x_i+E^{-1}_{ij}b_j$ and $\tilde{y}_i=y_i+E^{-1}_{ij}b_j$. Note that the factor $1/2$ is absent at the kinetic term in the Hamilton function \eqref{Ham_func}. Therefore, we have to stretch the proper-time and the potential in \eqref{model_simpl} accordingly in order to get \eqref{Ham_func} and \eqref{Ham_Jac_action}. Also notice that the Hamilton-Jacobi action \eqref{Ham_Jac_action} contains the inverse matrix $E^{-1}_{ij}$, but it has a finite limit when $E_{ij}$ becomes degenerate as seen from \eqref{Ham_Jac_action} expanded in $s$:
\begin{multline}\label{Ham_Jac_expan}
    S=\frac{(x-y)^2}{4s}+\frac12xfy+s\Bigl[\frac1{12}(x-y)f^2(x-y)+\frac16(xEx+yEy+xEy)+\frac12b(x+y)+E_0\Bigl]-\\
    -\frac{s^2}6\tilde{x}Ef\tilde{y}
    -\frac{s^3}{12}\Bigl[\frac1{15}(x-y)(f^2+2E)^2(x-y)+\tilde{x}E^2\tilde{y}\Bigr]+\ldots
\end{multline}
All the terms with $E^{-1}_{ij}$ cancel out. The higher terms of the expansion are independent of $E^{-1}_{ij}$. The van Vleck determinant is written as
\begin{equation}\label{vanVle}
\begin{split}
    \det\frac{\partial^2S}{\partial x^i\partial y^j}&=(-2s)^{-d}\det\frac{s\ka e^{sf}}{\sin s\ka}=(-2s)^{-d} \det\Bigl(\frac{\sin s\ka}{s\ka}\Bigr)^{-1},\\
    -\frac12\ln\frac{\sin s\ka}{s\ka}&=-\frac{s^2}{12}(f^2+2E)+\frac{s^4}{360}(f^2+2E)^2+\ldots,
\end{split}
\end{equation}
where we have used the unimodular property of the matrix $e^{sf}$.

Now we return to our problem. The leading (Gaussian) contribution to the low energy expansion of the heat kernel is obtained \cite{Schwing} if we expand the Hamiltonian \eqref{Hamiltonian} in the momentum and coordinate operators up to the second order and throw away the higher terms. Then we solve the Heisenberg equations for the retained quadratic Hamiltonian and find the evolution operator (the heat kernel) in the Fock proper-time \cite{Fock,Feyn,Schwing}. As well-known, this problem is essentially classical for the quadratic Hamiltonians, and the quasiclassical answer,
\begin{equation}\label{Green_func}
    \tilde{G}(\omega,s;\spx,\spy)=\lan\spx|e^{-is[-\tilde{H}(\omega)]}|\spy\ran=\Bigl[(-2\pi i)^{-d}\det\frac{\partial^2 S}{\partial x^i\partial y^j}\Bigr]^{1/2}e^{iS(s;x,y)},
\end{equation}
for the Green function is the exact one. Here tilde reminds us that the Green function is the bi-density rather than the bi-scalar as in Eq. \eqref{pos_freq_func}. Formula \eqref{Green_func} can also be obtained from the path-integral representation of the heat kernel after a Gaussian integration over the fields $x^i(\tau)$ (see, e.g., \cite{DeWGAQFT,TerZhukBor,BekPark}).

The Hamiltonian for our system is \eqref{Hamiltonian}, which we write as
\begin{multline}\label{Hamiltonian2}
%\begin{split}
    -\tilde{H}=\bar{g}^{-1/4}(p_i+\omega g_i)\sqrt{\bar{g}}\bar{g}^{ij}(p_j+\omega g_j)\bar{g}^{-1/4}+\frac12\bnabla^ih_i+\frac14h_ih^i+\frac{\omega^2}{\xi^2}+m^2=\\
    =(p_i+\omega g_i)\bar{g}^{ij}(p_j+\omega g_j)+\frac12\partial_i(\bar{g}^{ij}\partial_j\ln\sqrt{\bar{g}})+\frac14\partial_i\ln\sqrt{\bar{g}}\bar{g}^{ij}\partial_j\ln\sqrt{\bar{g}}+\frac12\bnabla^ih_i+\frac14h_ih^i+\frac{\omega^2}{\xi^2}+m^2.
%\end{split}
\end{multline}
Of course, we want to derive such an approximate expression for the heat kernel associated with this Hamiltonian that preserve all the symmetries of the exact evolution operator: it should be a kernel of a unitary operator, it should be a bi-density function which is invariant with respect to the gradient transformations of the field $g_i$. The last property guaranties that the corresponding positive-frequency function will be generally covariant under the spacetime transformations (see \cite{LandLifshCTF}, Sec. 88). Besides, we want that already the leading (Gaussian) approximation will give us the most exact approximation that we can achieve for the heat kernel in the case of the slowly varying fields $\bar{g}^{ij}$ and $g_i$, so as we need not to evaluate the higher order corrections to it using the perturbation theory, or reduce this work to a minimum.

%the higher order corrections to it, which can be evaluated using the perturbation theory, would be small

To this aim, we adopt the following strategy. For any given points $\spx$ and $\spy$ connected by the geodesic, we pass to the Riemann normal coordinates with the origin at the midpoint $p(\spx,\spy)$. In this frame, we expand the functions entering the Hamiltonian \eqref{Hamiltonian2} in a Taylor series. This automatically gives us the covariant expressions at any finite order of the expansion. In particular, using the formulas presented in \cite{Petrov,BekPark} and in Appendix \ref{Some_Expan}, we find
\begin{equation}\label{grav_pot}
\begin{split}
    \frac12&\partial_i(\bar{g}^{ij}\partial_j\ln\sqrt{\bar{g}})+\frac14\partial_i\ln\sqrt{\bar{g}}\bar{g}^{ij}\partial_j\ln\sqrt{\bar{g}}=-\frac16\bar{R}-\frac16\bnabla_i\bar{R}x^i+ \frac12\bar{r}_{ij}x^ix^j+\ldots,\\
    \bar{r}_{ij}&:=
    \frac1{5}\bigl(\frac13\bar{R}_{ik}\bar{R}^k_{\ j}-\frac16\bar{R}^{kl}\bar{R}_{kilj}-\frac16\bar{R}_i^{\ mnk}\bar{R}_{jmnk}-\frac14\bnabla^2\bar{R}_{ij}-\frac34\bnabla_{ij}\bar{R}\bigr).
\end{split}
\end{equation}
In passing from the bi-density to the bi-scalar, one should multiply the former by
\begin{equation}
    \bar{g}^{-1/4}(\spx)\bar{g}^{-1/4}(\spy)=\bar{\De}^{1/2}(\spx,\spy),
\end{equation}
where we have used the relation \cite{BekPark}
\begin{equation}
    \bar{\De}(\spx,0)=\bar{\De}(0,\spx)=\bar{g}^{-1/2}(\spx),
\end{equation}
and the composition property of the covariant van Vleck determinant. Also we need the expansion
\begin{equation}\label{Eij}
    \omega^2\xi^{-2}(x)=\omega^2\xi^{-2}\bigl(1-2h_ix^i+(2h_ih_j-\bnabla_ih_j)x^ix^j+\ldots\bigr)=:\omega^2(\xi^{-2}-b^{\br0}_ix^i-\frac12E^{\br0}_{ij}x^ix^j).
\end{equation}
Hereinafter, all the fields and their derivatives are assumed to be taken at the point $p$, unless otherwise stated. The midpoint prescription implies $\bs^i=2x^i=-2y^i$.

Now, in the Hamiltonian \eqref{Hamiltonian2}, we retain the terms which are at most quadratic in $x$ and $p$ (keeping the order they are written in \eqref{Hamiltonian2}) and obtain the Gaussian approximation for the heat kernel
\begin{equation}\label{HK_resum}
\begin{split}
    G(\omega,s;\spx,\spy)&=\frac{\bar{\De}^{1/2}(\spx,\spy)}{(4\pi i s)^{d/2}}\det\Bigl(\frac{\sin s\ka}{s\ka}\Bigr)^{-1/2}e^{iS_0(s,\bs_i)-isM^2},\\
    M^2:&=m^2-\frac16\bar{R}+\frac12\bnabla^ih_i+\frac14h^2=m^2-\frac16(R+\xi^2R_{\mu\nu}g^\mu g^\nu)-\frac14h^2,\\
    S_0:&=\frac14\bs\ka\ctg(s\ka)\bs-\frac12(\frac{\bs}2-bE^{-1})\ka\bigl(\ctg s\ka-\frac{e^{s\omega f}}{\sin s\ka}\bigr)(\frac{\bs}2+E^{-1}b)-\frac{s}2bE^{-1}b+\\
    &+\frac{\omega}2bE^{-1}f\bs-s\frac{\omega^2}{\xi^2},
\end{split}
\end{equation}
where, for brevity, we omit the matrix indices, $\bs\equiv\bs_i$ in what follows, $\ka$ is defined as above (see Eq. \eqref{Ham_Jac_action}) with the replacement $f\rightarrow\omega f$, and
\begin{equation}
\begin{gathered}
    E_{ij}=\omega^2E^{\br0}_{ij}+E^{\br2}_{ij},\qquad b_i=\omega^2 b^{\br0}_i+b^{\br2}_i,\\
    E^{\br2}_{ij}=\bnabla_{ij}\Bigl(\frac14h^2-\frac12\nabla^\la h_\la\Bigr)-\bar{r}_{ij},\qquad b^{\br2}_i=\bnabla_i\Bigl[\frac16(R+\xi^2R_{\mu\nu}g^\mu g^\nu)+\frac14h^2 \Bigr].
\end{gathered}
\end{equation}
Expanding the expression for $G$ in the number of derivatives, as it was done in the previous section, one can verify, using formulas \eqref{Ham_Jac_expan}, \eqref{vanVle}, that all the terms of the expansion \eqref{HK_qp_expand2} are reproduced save the term
\begin{equation}\label{higher_der_term}
    \frac{\omega s}6\bnabla^jf_{ji}\bs^i=\frac{\omega s}6(\nabla^\mu f_{\mu\nu}\bs^\nu+\bs^\nu f_{\nu\mu}h^\mu).
\end{equation}
It seems this term cannot be obtained from the Gaussian approximation. Such a term may come from the divergence of the gauge field $g_i$ that appears when one uses the $qp$-ordering of the Hamiltonian \eqref{Hamiltonian2}. However, the $qp$-ordered Hamiltonian truncated to its quadratic part is not Hermitian and the evolution operator is not unitary -- the property which we desire to preserve. Thus, substituting the expression obtained for $G$ to \eqref{pos_freq_func}, we arrive at the central result of the present paper
\begin{equation}\label{pos_freq_func_1}
    D^{(+)}(x,y)\approx-i\int ds\int_0^\infty\frac{d\omega}{2\pi}e^{-i\omega t_-}\frac{\bar{\De}^{1/2}(\spx,\spy)(1+\omega s\bnabla^jf_{ji}\bs^i/6)}{|\xi^2|^{1/4}(\spx)(4\pi i s)^{d/2}|\xi^2|^{1/4}(\spy)}\det\Bigl(\frac{\sin s\ka}{s\ka}\Bigr)^{-1/2}e^{iS_0(s,\bs_i)-isM^2},
\end{equation}
where $t\equiv T$ or the Fock gauge \eqref{Fock_gauge} based at the midpoint is implied. The higher derivative term \eqref{higher_der_term} could be ``exponentiated'', but we shall not investigate this possibility.

Some remarks are in order. Due to the noncommutativity of $x$ and $p$ in the kinetic part of the Hamiltonian \eqref{Hamiltonian2}, there is an ambiguity how to single out the quadratic part from it. For example, one could use the Weyl ordering, where the first term in \eqref{Hamiltonian2} without the gauge fields is rewritten as
\begin{equation}
    p_i\bar{g}^{ij}p_j=\frac14(p_ip_j\bar{g}^{ij}+2p_i\bar{g}^{ij}p_j+\bar{g}^{ij}p_ip_j)+\frac14\partial_{ij}\bar{g}^{ij}.
\end{equation}
Then one extracts the quadratic part from the Hamiltonian and obtains the same associated classical equations of motion as above, but with the additional correction to the potential. In particular, instead of $-\bar{R}/6$ in \eqref{grav_pot} one will have
\begin{equation}
    -\frac16\bar{R}+\frac14\partial_{ij}\bar{g}^{ij}(0)=-\frac14\bar{R}.
\end{equation}
It is this correction to the potential term which is argued by DeWitt as the correct one (see \cite{DeWGAQFT}, Chap. 15), when one uses the path-integral representation of the heat kernel with the midpoint prescription. However, with this potential term, one needs to perform the two-loop calculations of the path-integral in order to reproduce the well-known asymptotic of the heat kernel on the diagonal even at the first power of $s$. The potential with such a property is, of course, quite unsuitable for our purpose outlined above. In fact, changing the ordering prescriptions, one can obtain any number at the scalar curvature in the potential. The conditions, that we imposed on the approximate heat kernel, fix the prescription unambiguously. Moreover, it was proven in \cite{Parker} that the correction of the form \eqref{grav_pot} to the potential term sums all the terms of the derivative expansion of the heat kernel, which contain $\bar{R}$ at the given point. The estimates of the higher loop corrections are presented in Appendix \ref{Conf_Trans}. There we also discuss the conformal transformation technique applied to the heat kernel and how these transformations may change formulas \eqref{HK_resum} and \eqref{pos_freq_func_1}.

In order to provide a tighter connection of the positive-frequency function \eqref{pos_freq_func_1} with reality, let us write the quantities entering it in the weak field limit for the vacuum solutions of the Einstein equations (for details, see \cite{LandLifshCTF}, Sec. 105, and Appendix \ref{4_Kill})
\begin{equation}
\begin{gathered}
    b^{\br0}_i\approx-\frac{r_g}{r^2}n_i,\qquad b^{\br2}_i\approx-\frac{r_g}{4r^5}n_i,\qquad E^{\br0}_{ij}\approx-\frac{r_g}{r^3}(\de_{ij}-3n_in_j),\\
    \bar{r}_{ij}\approx-\frac{r_g^2}{80r^6}(5\de_{ij}-3n_in_j),\qquad E^{\br2}_{ij}\approx-\frac{3r_g^2}{16r^6}(\de_{ij}-\frac{39}{5}n_in_j),\\
    f_{ij}\approx\frac{2r_g}{mr^3}(M_{ij}+\frac32n_{[i}M_{j]k}n_k),\qquad
    [E,f]\approx\omega^2\frac{3r_g^2}{mr^6}n_{(i}M_{j)k}n_k,
\end{gathered}
\end{equation}
where $m$ is a total mass of the gravitating object, $M_{ij}$ is its angular momentum, $r_g$ is the Schwarzschild radius, $r$ is a distance from the gravitating object, and $n_i=x_i/r$. In the second line, the Schwarzschild metric was used for the calculations. We see that the tensor $E_{ij}$ has one positive and two negative eigenvalues as in \eqref{E_f}. The relation $[E,f]=0$, which we also assumed, is fulfilled when the vector of the angular momentum is parallel to the vector of the gravielectric force (see Appendix \ref{4_Kill} for the definition). The above calculations can be easily generalized to the case where the tensor $f_{ij}$ has a canonical form \eqref{E_f} and the tensor $E_{ij}$ is diagonal in the same basis. One should set $\la_1\in\mathbb{C}$ in the solution \eqref{eqm_sol}, and the matrix $e^{sf}$ in the Hamilton-Jacobi action \eqref{Ham_Jac_action} is to be placed near $\tilde{x}$. Nevertheless, further we restrict ourself to the case $[E,f]=0$.

%\newpage
\section{Physical implications}\label{Phys_Implic}

In this section we shall discuss some physical implications of the result obtained above. The crucial point is, of course, the fact that the positive-frequency function \eqref{pos_freq_func_1} depends nontrivially on the Killing vector. This dependence stems from the dependence of the Hamiltonian of the scalar field on the Killing vector of the gravitational background. The positive-frequency function is an observable and so it is just a matter of physical erudition to extract the dependence on the Killing vector from it. We shall find the various expressions for the propagator of the scalar field which are valid in the different regimes: at sufficiently large and sufficiently small point separations. Besides, we shall derive the exponentially suppressed contributions to the one-loop effective action which depend nontrivially on the vector field $\xi^\mu$. This action is induced by the polarization of the vacuum of scalar particles and represents the analog of the Heisenberg-Euler action \cite{HeisEul,Schwing} in quantum electrodynamics. In particular, we shall obtain the quasiclassical formulas for the Hawking particle production \cite{Hawk} analogous to Schwinger's formulas for the particle production in the constant electromagnetic field. The Unruh effect and the radiation reaction problem on a curved background will be also addressed and the formula for the acceleration determining the Unruh effect will be derived.

The structures appearing in the heat kernel \eqref{HK_resum} fall naturally into the pieces denoted by the indices $0$ and $2$. Let us estimate their ratio. Introducing the characteristic scales in the weak field limit
\begin{equation}\label{scales_weak}
    \partial_i\sim r^{-1}\sim L^{-1},\qquad h_i\sim\e/L,\qquad \bar{R}_{ij}\sim\e/L^2,\qquad \text{etc.}
\end{equation}
where $\e=(1-\xi^2)\sim r_g/L\ll1$, we see that the inequalities
\begin{equation}\label{assumptions}
    \omega^2|b^{\br0}_i|\gg|b^{\br2}_i|,\qquad \omega^2|E^{\br0}_{ij}|\gg|E^{\br2}_{ij}|,
\end{equation}
are equivalent to
\begin{equation}
    \omega^2L^2\gg1.
\end{equation}
Near the horizon, $\e\approx1$, the estimations \eqref{scales_weak} changes,
\begin{equation}\label{scales_strong}
    \partial_i\sim h_i\sim L^{-1}\xi^{-2},\qquad \bnabla_i h_j\sim\bar{R}_{ij}\sim L^{-2}\xi^{-4},\qquad\text{etc.}
\end{equation}
and the conditions \eqref{assumptions} become
\begin{equation}\label{cond_omega}
    \xi^4\omega^2L^2\gg1.
\end{equation}
The latter is valid in the weak field limit as well. Of course, certain combinations of the fields $g_i$, $f_{ij}$, $\bar{R}^i_{\ jkl}$, and their derivatives may be unexpectedly small or vanish on the particular solutions to the Einstein equations, but their orders of magnitude are not larger than given above. Loosely speaking, the inequality \eqref{cond_omega} says that we consider a propagation of the wave packet consisting of the modes with the wavelengths much smaller than the distance to the gravitating object. Therefore, in the range of applicability of the approximation of slowly varying fields, which we assume from the outset, the inequality \eqref{cond_omega} has to be hold and the estimations \eqref{assumptions} are fulfilled.

Now we evaluate the different types of contributions to the integral \eqref{pos_freq_func_1}. Stretching the proper-time $s\rightarrow s/\omega$ and redefining
\begin{equation}
    E_{ij}\rightarrow\omega^2E_{ij},\qquad b_i\rightarrow\omega^2b_i,
\end{equation}
we can write the integral in the form
\begin{multline}\label{pos_freq_func_2}
    D^{(+)}(x,y)=-i\int \frac{ds}{2\pi}\frac{\bar{\De}^{1/2}(\spx,\spy)(1+s\bnabla^jf_{ji}\bs^i/6)}{|\xi^2|^{1/4}(\spx)(4\pi i s)^{d/2}|\xi^2|^{1/4}(\spy)}\times\\
    \times\int_0^\infty d\omega\omega^{d/2-1}\det\Bigl(\frac{\sin s\ka}{s\ka}\Bigr)^{-1/2}e^{i\omega(S_0(s,\bs_i)-t_-)-isM^2/\omega},
\end{multline}
where $S_0$ is defined as in Eq. \eqref{HK_resum} with $\omega=1$. In the representation \eqref{E_f},
\begin{equation}\label{S_0}
\begin{split}
    S_0&=\frac{\bs_\perp^2}{8}\sqrt{H^2-2\la_1}\Bigl(\ctg s\sqrt{H^2-2\la_1}+\frac{\cos sH}{\sin s\sqrt{H^2-2\la_1}}\Bigr)+\frac{\bs_\parallel^2}8\sqrt{2\la_2}\tah\frac{s}{2}\sqrt{2\la_2}+\\
    &+\frac{b^2_\perp}{2\la_1^2}\sqrt{H^2-2\la_1}\Bigl(\ctg s\sqrt{H^2-2\la_1}-\frac{\cos sH}{\sin s\sqrt{H^2-2\la_1}}\Bigr)+\frac{b_\parallel^2}{2\la_2^2}\sqrt{2\la_2}\cth\frac{s}{2}\sqrt{2\la_2}-\\
    &-\frac1{2\la_1}(\bs_x b_y-\bs_y b_x)\Bigl(\sqrt{H^2-2\la_1}\frac{\sin sH}{\sin s\sqrt{H^2-2\la_1}}-H\Bigr)-s(\frac{b_\perp^2}{2\la_1}+\frac{b_\parallel^2}{2\la_2})-\frac{s}{\xi^2},
\end{split}
\end{equation}
and
\begin{equation}
    \det\Bigl(\frac{\sin s\ka}{s\ka}\Bigr)^{-1/2}=\frac{s\sqrt{H^2-2\la_1}}{\sin s\sqrt{H^2-2\la_1}}\Big(\frac{s\sqrt{2\la_2}}{\sh s\sqrt{2\la_2}}\Big)^{1/2},
\end{equation}
where $\ups_2$ is assumed to be directed along the $z$-axis, the projections to this axis are denoted as parallel, and the $x$ and $y$ projections are denoted as perpendicular.

\subsection{The $\omega$-representation}\label{Omega_Repres}

At first, we take the integral over $s$. The integrand has the three types of singular points in the $s$-plane: (a) the essentially singular points on the imaginary axis $s=i\pi n/\sqrt{2\la_2}$, (b) the essentially singular points on the real axis $s=\pi n/\sqrt{H^2-2\la_1}$, and (c) the branching point at the origin. The points of the type (a) are responsible for the Hawking particle production, the points of the type (b) describes the vacuum polarization effects, and the point (c) gives the major contribution to the propagator. None of the contributions from these points can be evaluated exactly. However we can make it approximately under the assumption that the estimation \eqref{cond_omega} holds. We shall expand $S_0$ near the singular points retaining only the leading terms of the Laurent series. The higher terms of this series give a negligible contribution. This is easy to see if one makes the variable $s$ dimensionless stretching it once more as $s\rightarrow s/\omega$.

\subsubsection{Exponentially suppressed contributions}

Introducing the notation
\begin{equation}
    l_{12}:=\frac{\sqrt{H^2-2\la_1}}{\sqrt{2\la_2}},\qquad l_{21}:=\frac{\sqrt{2\la_2}}{\sqrt{H^2-2\la_1}},\qquad l_{H1}:=\frac{H}{\sqrt{H^2-2\la_1}},\qquad l_{H2}:=\frac{H}{\sqrt{2\la_2}},
\end{equation}
the expansions near the points (a) can be cast into the form ($x\rightarrow0$)
\begin{equation}\label{S_0_odd}
\begin{split}
    S_0&\approx\frac{\bs_\parallel^2}{4x}-i\Bigl[\frac{\bs_\perp^2}{8}\sqrt{H^2-2\la_1}\Bigl(\cth\pi nl_{12}+\frac{\ch\pi n l_{H2}}{\sh\pi nl_{12}}\Bigr)
    +\frac{b_\perp^2}{2\la_1^2}\sqrt{H^2-2\la_1}\Bigl(\cth\pi nl_{12}-\frac{\ch\pi n l_{H2}}{\sh\pi nl_{12}}\Bigr)-\\
    &-\frac{i}{2\la_1}(\bs_x b_y-\bs_y b_x)\Bigl(\sqrt{H^2-2\la_1}\frac{\sh\pi nl_{H2}}{\sh\pi nl_{12}}-H\Bigr)+
    \frac{\pi n}{\sqrt{2\la_2}}(\xi^{-2}+\frac12bE^{-1}b)\Bigr],
\end{split}
\end{equation}
where $n$ is an odd number, and
\begin{equation}\label{S_0_even}
\begin{split}
    S_0&\approx\frac{b_\parallel^2}{\la_2^2x}-i\Bigl[\frac{\bs_\perp^2}{8}\sqrt{H^2-2\la_1}\Bigl(\cth\pi nl_{12}+\frac{\ch\pi n l_{H2}}{\sh\pi nl_{12}}\Bigr)
    +\frac{b_\perp^2}{2\la_1^2}\sqrt{H^2-2\la_1}\Bigl(\cth\pi nl_{12}-\frac{\ch\pi n l_{H2}}{\sh\pi nl_{12}}\Bigr)-\\
    &-\frac{i}{2\la_1}(\bs_x b_y-\bs_y b_x)\Bigl(\sqrt{H^2-2\la_1}\frac{\sh\pi nl_{H2}}{\sh\pi nl_{12}}-H\Bigr)+
    \frac{\pi n}{\sqrt{2\la_2}}(\xi^{-2}+\frac12bE^{-1}b)\Bigr],
\end{split}
\end{equation}
when $n$ is an even number. The mass term and the preexponential factor become in these cases
\begin{equation}\label{sm2}
    sM^2\approx\frac{i\pi nM^2}{\sqrt{2\la_2}},\qquad s^{-3/2}\det\Bigl(\frac{\sin s\ka}{s\ka}\Bigr)^{-1/2}\approx\frac{e^{-i\pi(n+1)/2}\sqrt{H^2-2\la_1}(2\la_2)^{1/4}}{\sh\pi nl_{12}\sh^{1/2}(x\sqrt{2\la_2})}.
\end{equation}
Recall that the integration contour in the $s$-plane goes along the real axis below the singularities lying on it. The half-plane (upper or lower), where we should close the contour, can be determined analyzing the asymptotic behavior of the integrand of \eqref{pos_freq_func_2} at the points (a) with large $n$. From formulas \eqref{S_0_odd}, \eqref{S_0_even}, and \eqref{sm2} we see that in the case when
\begin{equation}\label{omega_prop}
    \sqrt{H^2-2\la_1}>\omega(\xi^{-2}+\frac12bE^{-1}b)+M^2/\omega,
\end{equation}
the contour can be closed in the upper half-plane. Otherwise, we have to close the contour in the lower half-plane and the singular points of the types (b) and (c) do not contribute to the positive-frequency function at such $\omega$'s. One may say that such modes do not propagate.

For the energies $\omega$, which are much larger than the Compton wavelength and satisfy the estimation \eqref{cond_omega}, the inequality \eqref{omega_prop} is fulfilled. Then the contributions of the singularities (a) are suppressed by the Boltzmann-like factor (see Eqs. \eqref{S_0_odd}, \eqref{S_0_even})
\begin{equation}
    \exp[\omega\frac{\pi n}{\sqrt{2\la_2}}(\xi^{-2}+\frac12bE^{-1}b)]\approx\exp\Big[\omega\frac{\pi n}{\sqrt{2\la^{\br0}_2}}(\xi^{-2}+\frac12b^{\br0}E^{-1}_{\br0}b^{\br0})\Big],
\end{equation}
where $\la_2^{\br0}$ is the eigenvalue of the matrix $E^{\br0}$. The coefficient at the energy $\omega$ at $n=1$ can be interpreted as the reciprocal temperature of the Hawking radiation. Inasmuch as the one-loop contribution of one bosonic mode to the effective action reads as
\begin{equation}\label{one_loop}
    \Ga^{\br1}_{1b}=-i\int dx\frac{\sqrt{|g|}}{|\xi^2|}\int_0^\infty d\omega \omega^2D^{\br+}(\omega,\spx,\spx),
\end{equation}
these singularities contribute to the imaginary part of the effective action and suppressed by the same exponential factor. Formula \eqref{one_loop} is valid for a static metric. As for the general stationary case, see further Eq. \eqref{one_loop_2}. The spatial measure in the integral \eqref{one_loop} is that measure with respect to which the mode functions on the static spacetime are orthogonal. To evaluate the integral over $s$, we use the formula
\begin{equation}
    \int_H\frac{dz}{\sh^{1/2}(z\sqrt{2\la_2})}=-\frac{4\sqrt{2\pi}\Ga(5/4)}{\sqrt{2\la_2}\Ga(3/4)},
\end{equation}
where $H$ is the Hankel contour that runs from $+\infty$ a little bit higher than the real axis, encircles the origin, and then goes to $+\infty$ a little bit lower than the real axis. With this integral at hand, we can write the contributions from the singularities of the type (a) for odd $n$ ($n=1$ is the leading contribution) in the form
\begin{multline}\label{D_a}
    D^{\br+}(\omega,\spx,\spx)|_{\br{a}}\approx-ie^{-i\pi(n/2+1/4)}\frac{(2\la_2)^{1/4}\Gamma(5/4)}{\sqrt{2}\pi|\xi^2|^{1/2}\Ga(3/4)}\frac{l_{12}}{\sh\pi nl_{12}}\times\\
    \times\omega^{1/2} e^{\omega\frac{b_\perp^2}{2\la_1^2}\sqrt{H^2-2\la_1}\bigl(\cth\pi nl_{12}-\frac{\ch\pi n l_{H2}}{\sh\pi nl_{12}}\bigr)+\frac{\pi n}{\sqrt{2\la_2}}\big(\frac{\omega}{\xi^2}+\frac{\omega}2bE^{-1}b-\frac{M^2}{\omega}\big)}.
\end{multline}
The imaginary part of the correction to the effective action coming from these singularities is positive as it should be. There exists also the exponentially suppressed contribution to the real part of the effective action from these terms. As far as the contributions with even $n$ are concerned, the nonvanishing term at $1/x$ in the expansion \eqref{S_0_even} leads to a complication in evaluating the integral over $s$ and the resulting expression is rather huge. So, we do not write it here.

Near the points (b), we have
\begin{equation}\label{S_0_magn}
\begin{split}
    S_0&\approx\frac1{x}\Bigl[\frac{\bs_\perp^2}{8}(1+(-1)^n\cos\pi nl_{H1})+\frac{b^2_\perp}{2\la_1^2}(1-(-1)^n\cos\pi nl_{H1})
    -\frac{(-1)^n}{2\la_1}(\bs_x b_y-\bs_y b_x)\sin\pi nl_{H1}\Bigr]-\\
    &-H(-1)^n\sin\pi nl_{H1}(\frac{\bs_\perp^2}{8}-\frac{b^2_\perp}{2\la_1^2})
    +\sqrt{2\la_2}\cth\pi nl_{21}(\frac{\bs_\parallel^2}{8}+\frac{b_\parallel^2}{2\la_2^2})+\\
    &+\frac{H}{2\la_1}(\bs_x b_y-\bs_y b_x)(1-(-1)^n\cos\pi nl_{H1})
    -\frac{\pi n}{\sqrt{H^2-2\la_1}}(\xi^{-2}+\frac12bE^{-1}b),
\end{split}
\end{equation}
and
\begin{equation}
    sM^2\approx\frac{\pi nM^2}{\sqrt{H^2-2\la_1}},\qquad s^{-3/2}\det\Bigl(\frac{\sin s\ka}{s\ka}\Bigr)^{-1/2}\approx\frac{(-1)^n(2\la_2)^{1/4}}{x\sh^{1/2}(\pi nl_{21})},
\end{equation}
where $x\rightarrow0$. The integrand of \eqref{pos_freq_func_2} is a single-valued function in the neighbourhood of these singular points because of the degeneracy of the eigenvalue $\sqrt{H^2-2\la_1}$. As a result, the integral over $s$ going around these points can be simply evaluated. Assuming the inequality \eqref{omega_prop} is satisfied, we arrive at
\begin{equation}\label{D_b}
    D^{\br+}(\omega,\spx,\spy)|_{\br{b}}\approx \frac{\bar{\De}^{1/2}(\spx,\spy)\Big(1+\frac{\pi n\bnabla^jf_{ji}\bs^i}{6\sqrt{H^2-2\la_1}}\Big)(-1)^n(2\la_2)^{1/4}}{|\xi^2|^{1/4}(\spx)(4\pi i )^{3/2}|\xi^2|^{1/4}(\spy)\sh^{1/2}(\pi nl_{21})}
    \omega^{1/2}e^{i\omega S^f_0-\frac{i\pi nM^2}{\omega\sqrt{H^2-2\la_1}}},
\end{equation}
where $S_0^f$ is the finite part of the expansion \eqref{S_0_magn} at $x\rightarrow0$.

\subsubsection{Main contribution}

The main contribution to the positive-frequency function comes from the origin of the $s$-plane. From the expansion \eqref{Ham_Jac_expan} near this point taken with the midpoint prescription, we deduce
\begin{equation}\label{S_0_expan}
    S_0=\frac{\bs^2}{4s}+s\big[\frac1{24}\bs(2f^2+E)\bs-\frac{1}{\xi^2}\big]+\frac{s^2}{6}\bs fb-\frac{s^3}{12}\big[b^2+\frac1{15}\bs(f^2+2E)^2\bs-\frac14\bs E^2\bs\big]+\ldots
\end{equation}
Rescaling the proper-time $s\rightarrow s/\omega$ and bearing in mind that $S_0$ is multiplied by $\omega$ in the exponent \eqref{pos_freq_func_2}, we see that the terms of the above expansion at the second power of $s$ and higher are small provided the condition \eqref{cond_omega} holds. Besides, there are also the saddle points near (after the rescaling)
\begin{equation}\label{extrem}
    s^2=\frac{\omega^2\bs^2}{4(\tilde{E}-\tilde{m}^2/\omega^2)},\qquad \tilde{E}:=-1/\xi^2+\bs(2f^2+E^{\br0})\bs/24,\qquad \tilde{m}^2:=M^2-\bs E^{\br2}\bs/24.
\end{equation}
If these points are situated near the origin they will considerably contribute to the integral and we cannot evaluate it simply by expanding $S_0$ near $s=0$. Here we are interested in the approximate expression for the propagator at the point separation much larger than the wavelength of a mode (for the case of the infinitely small $\bs$ and $t$ see the next subsection). So, we assume that the extremum points are far from the singular point $s=0$. This condition is fulfilled when
\begin{equation}\label{large_sep}
    |\xi^2|\omega^2\bs^2\gg1.
\end{equation}
Hence, keeping only the terms at $s^{-1}$, $s^0$, and $s$ in the exponent of \eqref{pos_freq_func_2} and the leading contribution from the preexponential factor, we obtain the approximate expression
\begin{equation}\label{M_eff}
\begin{split}
    D^{(+)}(x,y)|_{\br{c}}\approx\frac{-i\bar{\De}^{1/2}(\spx,\spy)}{|\xi^2|^{1/4}(\spx)|\xi^2|^{1/4}(\spy)}\int \frac{ds}{2\pi}
    \int_0^\infty d\omega\frac{\omega^{d/2-1}}{(4\pi i s)^{d/2}} e^{i\omega(\frac{\bs^2}{4s}-t_-+s\tilde{E})-is\tilde{m}^2/\omega}.
\end{split}
\end{equation}
The integral over $s$ is reduced to the Bessel function $J_{1/2}$ for $d=3$ and, consequently, is expressed in terms of elementary functions. After a little algebra, we find
\begin{equation}\label{pos_freq_func_main_omega}
    D^{(+)}(x,y)|_{\br{c}}=\frac{-i\bar{\De}^{1/2}(\spx,\spy)}{|\xi^2|^{1/4}(\spx)|\xi^2|^{1/4}(\spy)}
    \int_0^\infty \frac{d\omega}{4\pi^2|\bs|}\theta(\omega^2\tilde{E}-\tilde{m}^2)e^{-i\omega t_-}\sin(|\bs|\sqrt{\omega^2\tilde{E}-\tilde{m}^2}).
\end{equation}
The latter integral is the same as the integral for the positive-frequency function in a flat spacetime at $\tilde{m}^2\geq0$ with the obvious redefinitions. Applying the formulas from the Appendix of \cite{BSh}, we come to ($\tilde{m}^2\geq0$)
\begin{equation}\label{pos_freq_func_main}
\begin{split}
    D^{(+)}|_{\br{c}}&=\frac{-\bar{\De}^{1/2}(\spx,\spy)}{|\xi^2|^{1/4}(\spx)\tilde{E}^{1/2}|\xi^2|^{1/4}(\spy)}\bigg\{\frac{\sgn(t)}{4\pi}\de(\la)+\\ &+\frac{i\tilde{m}\theta(\la)}{8\pi\sqrt{\la}}\big[N_1(\tilde{m}\sqrt{\la})+i\sgn(t)J_1(\tilde{m}\sqrt{\la}) \big]+\frac{i\tilde{m}\theta(-\la)}{4\pi^2\sqrt{-\la}}K_1(\tilde{m}\sqrt{-\la})\bigg\}\approx\\
    &\approx\frac{-\bar{\De}^{1/2}(\spx,\spy)}{|\xi^2|^{1/4}(\spx)\tilde{E}^{1/2}|\xi^2|^{1/4}(\spy)}\Big[\frac{\sgn(t)}{4\pi}\de(\la)-\frac{i}{4\pi^2\la}+\frac{i\tilde{m}^2}{8\pi^2} \ln\frac{\tilde{m}|\la|^{1/2}}{2}-\frac{\tilde{m}^2}{16\pi}\sgn(t)\theta(\la)\Big],\\
    \la&\equiv-\tilde{X}^2(x,y):=\tilde{E}^{-1}t^2-\bs^2,
\end{split}
\end{equation}
where the last approximate expression is the expansion of the positive-frequency function at the small effective mass $\tilde{m}$. Of course, the positive-frequency function obtained can be written in a more compact form with the $i\e$-prescription $t\rightarrow t_-$. The unfolded form presented in \eqref{pos_freq_func_main} allows us to see better the structure of its singularities. It is clear that all the other Green functions can be obtained from the positive-frequency function.

Such an expression for the main contribution to the positive-frequency function has several physical implications. First, we see that the wave packet of scalar particles of the mass $m$ in a slowly varying gravitational field behaves like a massive particle with the effective mass $\tilde{m}$ given in \eqref{extrem}. In particular, if we neglect the term in $\tilde{m}^2$ proportional to $\bs^2$ and consider the vacuum solution to the Einstein equations then the mass squared acquires the shift which is equal to $-h^2/4$ (for an analogous but not the same effect in quantum electrodynamics see \cite{Ritus2p}). This is a tiny negative quantity and is of the order of the Unruh temperature squared. One could expect the appearance of such a correction to the mass already from the expression \eqref{Hamiltonian} of the Fourier transformed Klein-Gordon operator. Notice that this correction to the mass squared is also necessary to reproduce the standard asymptotic expansion of the heat kernel \eqref{HK_stand} and the propagator (see Eq. \eqref{pos_freq_func_s_main} below). A reliability of this correction against the higher loop corrections is considered in Appendix \ref{Conf_Trans}, where it is shown that this correction is not overridden by the higher loop corrections, at least for a strong field $\xi^2\approx0$. This correction makes no trouble for the massive particles since $h^2$ is small, but for the massless scalar particles it seems result in the tachyonic dispersion law in the adapted system of coordinates and to the instability of a vacuum. If $\tilde{m}^2<0$ and small, the additional correction appears in the positive-frequency function
\begin{equation}
    \de D^{(+)}(x,y)|_{\br{c}}\approx\frac{-\bar{\De}^{1/2}(\spx,\spy)}{|\xi^2|^{1/4}(\spx)\tilde{E}^{1/2}|\xi^2|^{1/4}(\spy)}\frac{\tilde{m}^2}{4\pi^2},
\end{equation}
which comes from the accurate evaluation of the integral \eqref{pos_freq_func_main_omega} and should be added to the expansion in the third line of \eqref{pos_freq_func_main}.

Certainly, we should not regard the expression \eqref{pos_freq_func_main} too seriously for the modes with the energies squared much less than $h^2$ where the considerable tachyonic effects may appear. These modes are out of the range of the applicability of the approximation we made in deriving \eqref{pos_freq_func_main}. However, for the massive particles and for the high energy modes of the massless ones satisfying \eqref{cond_omega}, \eqref{large_sep} the positive-frequency function in the representations \eqref{pos_freq_func_main_omega} or \eqref{pos_freq_func_main} with the corrections \eqref{D_a} and \eqref{D_b} describes a propagation quite well. The negative mass squared manifests as the vacuum instability and leads to the Bose-Einstein condensation. If one has in view that the gravitational field produces particles at the Hawking temperature then the appearance of such a correction to mass becomes expectable. Similar corrections to the mass squared are well-known in quantum field theory at finite temperature \cite{GrPeYa,DolJack,Kapusta}. Also note that the presence of matter can lead to the negative correction to the mass squared according to formula \eqref{HK_resum}. This is the well-known Jeans instability.

Second, the effective interval $\tilde{X}^2$ does not coincide with the interval \eqref{quasiinterval}. Indeed, it is easy to see for the worldline of a particle staying at rest, $\bs=0$, that the equation $X^2(\tau)=0$ has the nontrivial complex solutions (see Eqs. \eqref{quasiinterval}, \eqref{quasiint_exp}), while the equation $\tilde{X}^2(\tau)=0$ has not. If we assume
\begin{equation}
    |\xi^{-2}|\gg|\bs(2f^2+E^{\br0})\bs/24|,
\end{equation}
then
\begin{equation}\label{effinterval_0}
    \tilde{X}^2\approx\xi^2t^2+\frac{\xi^4t^2}{24}\bs(2f^2+E^{\br0})\bs+\bs^2.
\end{equation}
The latter expression differs from \eqref{quasiinterval} only by the terms in the square brackets in \eqref{quasiinterval} which contain $T$. It readily follows from this observation the expansion of $\tilde{X}^2$ on the arbitrary worldline:
\begin{equation}\label{effinterval}
    \tilde{X}^2\approx-\tau^2-\frac1{12}a^2_{U}\tau^4+\ldots,\qquad a^2_{U}:=(\nabla_uu)^2-[(a^{gm}_\mu+a^{ge}_\mu)^2-a_{gm}^2],
\end{equation}
where $a^{gm}_\mu:=(\xi u)f_{\mu\nu}u^\mu$ and $a^{ge}_\mu:=(\xi u)(gu)h_\mu$ are the gravimagnetic and gravielectric forces divided by the mass of a particle, respectively (see Appendix \ref{4_Kill} for details). In particular, the acceleration squared $a_U^2$ is zero for a resting particle when $u^\mu=\xi^\mu|\xi^2|^{-1/2}$. We shall call the acceleration $a_U$ as the Unruh acceleration since it determines the Unruh effect for a detector moving in a curved spacetime. For other attempts to construct similar quantity in the context of the radiation reaction problem see, e.g., \cite{Hiraya}.

As well known (see, e.g., \cite{DeWGAQFT,Unruh}), the transition rate of the detector to the excited state is determined by the Fourier transform of the positive-frequency function taken on the worldline of the detector. More precisely, for the two points $x(\tau_1)$ and $x(\tau_2)$ we introduce $\bar{\tau}:=(\tau_1+\tau_2)/2$ and $\tau:=\tau_1-\tau_2$, and write $D^{\br+}(x(\tau_1),x(\tau_2))=D^{\br+}(\bar{\tau},\tau)$. Then the rate of transition of the detector to the excited state $\omega>0$ is proportional to
\begin{equation}\label{trans_rate}
    \int d\tau e^{-i\omega\tau}D^{\br+}(\bar{\tau},\tau),
\end{equation}
where the integration contour in the $\tau$-plane passes slightly below the real axis. The positive-frequency function $D^{\br+}(\tau)$ is singular at $\tau=0$, where it possesses the pole of second order (for the non-isotropic worldline). All its other singularities are arranged symmetrically with respect to the imaginary axis. Therefore, up to the fourth order in $\tau$, they lie only on the real or imaginary axes when $a_U^2<0$ and $a_U^2>0$, respectively. In the former case, the Unruh detector feels nothing since we can close the integration contour in the lower half-plane. Of course, there may be singularities that are not catched by the expansion \eqref{effinterval}, but they lie far from the real axis and so they are suppressed. In the case $a_U^2>0$, the Unruh detector will detect the excitations at the reciprocal temperature
\begin{equation}\label{UNr_temp}
    \beta_U=\frac{2\pi}{a_U},
\end{equation}
where we have replaced the factor $\sqrt{12}$ by $2\pi$ matching the formula for the Unruh temperature with its flat spacetime analog for the hyperbolic motion. A concrete value of this factor depends on a character of the particle motion and requires a more detailed information about its worldline. Nevertheless, formula \eqref{UNr_temp} provides a good approximation for the temperature by the order of magnitude. Thus we see that the Unruh detector can be employed to test the variations of the field $\xi^\mu$. Notice that if we used the interval \eqref{quasiinterval} in order to describe the Unruh effect, we would obtain $(\nabla_uu)^2$ for the Unruh acceleration and, consequently, arrive at the unphysical result that the resting Unruh detector gets excited in a stationary gravitational field. It should be stressed the important difference between the Hawking \cite{Hawk} and the Fulling-Unruh \cite{Fulling,Unruh} effects. The Hawking particle production is caused by the action of the gravitational forces, and the detector at rest may records this process. Contrarily, the Unruh detector responds to the action of the inertial forces. The latter can be defined as a $4$-vector (see Appendix \ref{4_Kill}) as long as the vector field $\xi^\mu$ exists on the manifold.

The fact that the effective interval \eqref{effinterval_0} differs from the geodetic interval results also in a nonvanishing local expression for the radiation reaction force acting on the charged particle in a free fall. It is not difficult to show (see, e.g., \cite{DeWBr,Hobbs,Pois,GalSpir}) that the retarded Green function constructed from the propagator with the interval \eqref{quasiinterval} (see Eq. \eqref{pos_freq_func_s_main} below) yields the local contribution to the radiation reaction force proportional to
\begin{equation}
    \nabla_u\nabla_uu_\mu-(\nabla_uu)^2u_\mu,
\end{equation}
on the vacuum solutions to the Einstein equations. This contribution vanishes for the geodesic motion and is not zero for a particle at rest. As for the Green function with the effective interval \eqref{effinterval}, the situation will be reversed: the radiation reaction force acting on the particle at rest will be zero (though the ponderomotive force can be nonvanishing, see for a review \cite{Husnut}), and it will be nonzero for the particle in a free fall. In a certain sense, the additional terms appearing in the effective equations of motion of a particle and stemming from the difference of \eqref{pos_freq_func_main} from the DeWitt anzats take effectively into account the so-called tail term of the radiation reaction force. We postpone a thorough investigation of this problem to a future research, but should note that the charged particles can be also used to detect variations of the vector field $\xi^\mu$.

One may wonder why the retarded (advanced) Green function following from \eqref{pos_freq_func_main} does not satisfy the general theorems concerning the structure of singularities of the fundamental solution to the hyperbolic partial differential equation \cite{Hadamard,Cour}. The singularities of such a solution must lie on the characteristic cone determined by the geodetic interval, but we saw that the effective interval \eqref{effinterval_0} does not coincide with this interval. The answer is that the expression for the main contribution to the positive-frequency function \eqref{pos_freq_func_main} is valid only for the point separation much larger than the wavelengths of the modes in the wave packet (see Eq. \eqref{large_sep}). And so, the general theorem does not apply to the expression \eqref{pos_freq_func_main}. In order that the wave packet propagates along the characteristic surface, it should be delta-shaped in the space at the initial moment and consists of the ultrarelativistic modes. However, the delta-shaped wave packet is infinitely broad in the frequency space and does not obey the restriction \eqref{large_sep}. From the uncertainty relation for massless particles
\begin{equation}
    \De\omega\De x\geq2\pi,
\end{equation}
where $\De\omega$ is a width of the wave packet in the frequency space and $\De x$ is its width in the space, we find that the above expression \eqref{pos_freq_func_main} for the Green function holds only for the wave packets with
\begin{equation}
    \De x\gg\frac{2\pi}{\omega_0},
\end{equation}
where $\omega_0$ is the central frequency of the wave packet of the ultrarelativistic (massless) particles. Roughly speaking, the fact that the singular surface of the contribution \eqref{pos_freq_func_main} to the positive-frequency function deviates from the geodetic light cone just indicates that the wave packet with the extension much larger than the wavelengths of its modes moves slightly different than an ideal point massless particle. In the next subsection, we shall see that in the limit $(t,\bs)\rightarrow0$ (violating the condition \eqref{large_sep}) the structure of singularities of the positive-frequency function \eqref{pos_freq_func_2} is such as dictated by the theorems.

\subsection{The $s$-representation}\label{Propert_Repres}

Hitherto we have analyzed the $\omega$-representation of the positive-frequency function and derived, in particular, the expression for the major contribution to it using this representation. It turns out that under the assumption \eqref{cond_omega} we can take the integral over $\omega$ in \eqref{pos_freq_func_2} with a negligible error. Thereby we shall derive the $s$-representation of the positive-frequency function. This representation will be employed to obtain the standard asymptotic expansion of the propagator in terms of the geodetic interval squared $2\s$ at $(t,\bs)\rightarrow0$.

\begin{figure}[t]
\centering

\includegraphics*[width=15cm]{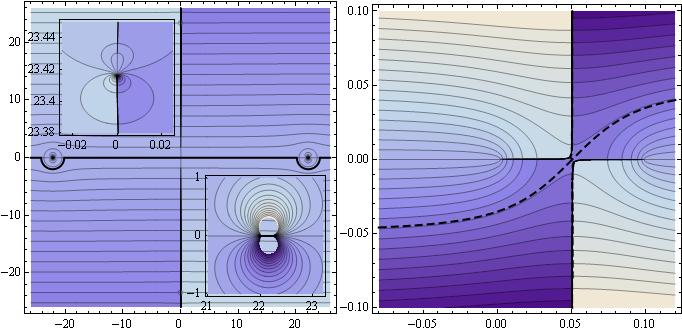}

\caption{{\footnotesize The typical contour plot of the imaginary part of the function standing in the argument of the Hankel function in Eq. \eqref{pos_freq_func_s}. The imaginary part changes its sign when reflected in the real axis. The insets depict the structure of singularities of the types (a) and (b) and the cuts near them.  The thin lines on all these plots are the lines of the steepest descent. On the left panel: The thick line going along the real axis is the initial integration contour. On the right panel: The structure of singularities near the origin is depicted. The dashed line shows the steepest descent contour. The ends of the cuts lying on the real axis are the zeroes of $a(s)$ nearest to the origin. The small line connecting two cuts intersects the integration contour at the saddle point.}}
\label{cuts}
\end{figure}

If the condition \eqref{cond_omega} is satisfied, we can expand $S_0$ standing in the exponent in formula \eqref{pos_freq_func_2} in the Laurent series in $\omega$ and retain only the terms at $\omega$ and $\omega^{-1}$. This can be easily done under the assumption that
\begin{equation}\label{commutativ}
    [E^{\br2},E^{\br0}]\approx0,\qquad [E^{\br2},f]\approx0.
\end{equation}
The latter relations are valid in the weak field limit as well as for the spherically symmetric metrics. Then $S_0$ in \eqref{pos_freq_func_2} keeps its form with
the replacements $E\rightarrow E^{\br0}$ and $b\rightarrow b^{\br0}$, while $M^2$ acquires the correction coming from the expansion of $S_0$. The expression for this correction is rather huge.
At the coincidence limit it reads as
\begin{multline}%\label{deM2}
    \de M^2|_{\bs=0}=b^{\br2}E_{\br0}^{-1}b^{\br0}-\frac12 b^{\br0}E_{\br0}^{-1} E_{\br2}E_{\br0}^{-1}b^{\br0}+\frac12 b^{\br0}E_{\br0}^{-1}\frac{E_{\br2}}{\sin^2s\ka}(\cos s\ka\ch sf-1)E_{\br0}^{-1}b^{\br0}-\\
    -\frac1{s}\Big[b^{\br2}E_{\br0}^{-1}\ka+b^{\br0}E_{\br0}^{-1}E_{\br2}\big(E_{\br0}^{-1}f^2+\frac32\big)\ka^{-1}\Big] \big(\ctg s\ka-\frac{\ch sf}{\sin s\ka}\big)E_{\br0}^{-1}b^{\br0},
\end{multline}
where $\ka$ is defined as before, but with $E\rightarrow E^{\br0}$. At the small $s$, this correction behaves as
\begin{equation}\label{deM21}
    \de M^2=-\frac{\bs E^{\br2}\bs}{24}+O(s\bs).
\end{equation}
This expression is valid even if Eqs. \eqref{commutativ} do not hold. The preexponential factor in \eqref{pos_freq_func_2} does not change apart from the redefinition of $\ka$ mentioned above. As a result, the integral over $\omega$ is reduced to the Hankel function \cite{GrRy}:
\begin{equation}\label{Hankel_func}
    \int_0^\infty d\omega\omega^{d/2-1}e^{i\omega(S_0-t_-)-is\bar{m}^2/\omega}=i^{d+1}\pi (is)^{d/2}\Bigl(\frac{\bar{m}^2}{a}\Bigr)^{d/4}H^{(1)}_{d/2}(2\bar{m}a^{1/2}),
\end{equation}
where $\bar{m}^2:=M^2+\de M^2$, the square root in the argument of the Hankel function has the cut along the positive real semi-axis, and $a:=s(t_--S_0)$,
\begin{multline}\label{a(s)}
    a=s^2(\xi^{-2}+\frac{1}2b_{\br0}E^{-1}_{\br0}b_{\br0})-s(t_-+\frac{1}2b_{\br0}E^{-1}_{\br0}f\bs)-\\
    -\bs\frac{s\ka}4\ctg s\ka\bs+(\frac{\bs}2-b_{\br0}E^{-1}_{\br0})\frac{s\ka}2\bigl(\ctg s\ka-\frac{e^{s f}}{\sin s\ka}\bigr)(\frac{\bs}2+E^{-1}_{\br0}b_{\br0}).
\end{multline}
In our case, $d=3$, the Hankel function is expressed in terms of the elementary functions
\begin{equation}\label{Hankel_func_3}
    \Bigl(\frac{\bar{m}^2}{a}\Bigr)^{d/4}H^{(1)}_{d/2}(2\bar{m}a^{1/2})=-\frac{i+2\bar{m}a^{1/2}}{2\pi^{1/2}a^{3/2}}e^{2i\bar{m}a^{1/2}}.
\end{equation}
Thus the positive-frequency function can be written as
\begin{equation}\label{pos_freq_func_s}
    D^{\br+}(x,y)=i^d\int\frac{ds \bar{\De}^{1/2}(\spx,\spy) (1+s\bnabla^jf_{ji}\bs^i/6)}{2|\xi^2|^{1/4}(\spx)(4\pi)^{d/2}|\xi^2|^{1/4}(\spy)}  \det\Bigl(\frac{\sin s\ka}{s\ka}\Bigr)^{-1/2} \Bigl(\frac{\bar{m}^2}{a}\Bigr)^{d/4}H^{(1)}_{d/2}(2\bar{m}a^{1/2}).
\end{equation}
The singularities of the integrand are located at the points (a) and (b) discussed in the previous subsection and also at the points where $a(s)=0$. The cuts stemming from the square roots appearing in the argument of the Hankel function and the determinant stretch between these singular points as depicted on Fig. \ref{cuts}. Recall that the integration contour in the $s$-plane lies a little bit lower than the real axis.

Unfortunately, the integral \eqref{pos_freq_func_s} cannot be evaluated exactly. Therefore we shall calculate it by the steepest descent method. To this end, we have to deform the integration contour and rise it from the fourth quadrant to the first one (see Fig. \ref{cuts}, the right panel). This gives the additional contributions to the integral from the cuts located on the positive real semi-axis, which are responsible for the vacuum polarization effects. We have already evaluated such contributions in the previous subsection and so here we concentrate on the major contribution to the positive-frequency function and its singularities.

The main contribution to the integral \eqref{pos_freq_func_s} comes from the extremum positioned near the point $s=0$. The singularities of the positive-frequency function appear when the two branching points nearest to the origin (zeroes of $a(s)$) approach the saddle point and pinch the integration contour. At $(t,\bs)\rightarrow0$, this saddle point and the two branching points tend to the origin $s=0$. In this limit the correction to $M^2$ is small (see Eq. \eqref{deM21}) and $\bar{m}$ can be regarded as a constant. Hence, these saddle and branching points are determined solely by the function $a(s)$. To apply the steepest descent method we introduce a new variable $z$ in the vicinity of the extremum such that
\begin{equation}\label{var_z}
    z^2+z_0^2=-a(s),
\end{equation}
where $z_0$ is a constant to be determined and $z=0$ corresponds to the extremum point $s=s_0$. Making use of the expansion \eqref{S_0_expan}, we have for \eqref{a(s)}:
\begin{equation}\label{a(s)_exp}
    a=-\frac{\bs^2}4+st-s^2\tilde{E}-\frac{s^3}{6}b^{\br0}f\bs+\frac{s^4}{12}\big[b^2_{\br0}+\frac1{15}\bs(f^2+2E^{\br0})^2\bs-\frac14\bs E^2_{\br0}\bs \big]+O(s^5),
\end{equation}
where we imply that $t$ has a small negative imaginary part. Now the extremum $s_0$ of the function $a(s)$ can be found perturbatively assuming that $t\sim\bs\sim l\rightarrow0$, where $l$ characterizes the point separation. Simple calculations give for the saddle point
\begin{equation}
    s_0=\frac{t}{2\tilde{E}}+\frac38\frac{c_3t^2}{\tilde{E}^3}+\big(\frac9{16}\frac{c_3^2}{\tilde{E}^5}+\frac{c_4}{4\tilde{E}^4}\big)t^3+\ldots,
\end{equation}
where $c_3$ and $c_4$ are the coefficients of the expansion of the function $a(s)$ at $s^3$ and $s^4$ in Eq. \eqref{a(s)_exp}, respectively. At the saddle point the function $a(s)$ is proportional to the geodetic interval squared
\begin{equation}
    a(s_0)=-\frac14\Big\{\bs^2+\xi^2t^2+\frac1{12}\xi^4t^2\Big[\bs^i\big(f^2_{ij}+\xi^{-2}(\bnabla_ih_j-h_ih_j)\big)\bs^j-2h^if_{ij}\bs^j t-h^2t^2\Big]\Big\}+\ldots=-\frac14\s_\mu\s^\mu.
\end{equation}
This determines the constant $z_0$ from \eqref{var_z}.

Carrying out the calculations, we shall keep the terms in \eqref{pos_freq_func_s} up to the second order in $l$ only. This will allow us to find the singular and finite parts of the positive-frequency function at $l\rightarrow0$. In particular, according to this approximation $\bar{m}$ should be replaced by $M$ as seen from \eqref{deM21} and \eqref{pos_freq_func_s}, since $s\sim z\sim l$ and
\begin{equation}
    \bar{m}a^{1/2}=Ma^{1/2}+O(l^3).
\end{equation}
Thus, differentiating \eqref{var_z}, we have
\begin{equation}
\begin{gathered}
    \dot{z}\approx\sqrt{\tilde{E}-3c_3s_0-6c_4s_0^2},\qquad\ddot{z}\approx-\frac{c_3+4c_4s_0}{\dot{z}},\qquad\dddot{z}\approx-\frac{3c_4}{\dot{z}}-\frac{3}{4\dot{z}^3}(c_3+4c_4s_0)^2,\\
    s\approx s_0+\frac{z}{\dot{z}}+\frac{c_3+4c_4s_0}{2\dot{z}^4}z^2+\big[c_4+\frac{7}{12}(c_3+4c_4s_0)^2\big]\frac{z^3}{2\dot{z}^5},
\end{gathered}
\end{equation}
where the overdots denote the derivatives with respect to $s$ taken at $s=s_0$. Introducing the notation $A_0$, $A_1$, and $A_2$ for the coefficients of the expansion in $s$ of the factor in \eqref{pos_freq_func_s} standing at the combination \eqref{Hankel_func_3}, we can write
\begin{multline}
  \frac{ds}{dz}(A_0+A_1s+A_2s^2)\approx\frac{A_0+A_1s_0+A_2s^2_0}{\dot{z}}+ \\
    +\big[A_1+2A_2s_0+(A_0+A_1s_0)\frac{c_3+4c_4s_0}{\dot{z}^2}\big]\frac{z}{\dot{z}^2} +(A_2+\frac{3c_4}{2\dot{z}^2}A_0)\frac{z^2}{\dot{z}^3}.
\end{multline}
The terms at the odd powers of $z$ can be omitted as long as the integration over $z$ is carried out on the symmetric interval. As a result, the integral in \eqref{pos_freq_func_s} can be cast into the form
\begin{multline}
    2\frac{\partial}{z_0\partial z_0}\int_0^\infty dz\Big[\frac{A_0+A_1s_0+A_2s^2_0}{\dot{z}}+(A_2+\frac{3c_4}{2\dot{z}^2}A_0)\frac{z^2}{\dot{z}^3}\Big]\frac{e^{-2M\sqrt{z^2+z_0^2}}}{\sqrt{z^2+z_0^2}}=\\
    =\frac{4M}{\dot{z}z_0}(A_0+A_1s_0+A_2s^2_0)K_1(2Mz_0)+\frac{2}{\dot{z}^3}(A_2+\frac{3c_4}{2\dot{z}^2}A_0)K_0(2Mz_0).
\end{multline}
Using the explicit formulas for $A_i$
\begin{equation}
\begin{split}
    A_0&=\frac{-i}{32\pi^2}\frac{\bar{\De}^{1/2}(\spx,\spy)}{|\xi^2|^{1/4}(\spx)|\xi^2|^{1/4}(\spy)}\approx\frac{-i}{32\pi^2|\xi^2|^{1/2}(p)}[1-\frac18\bnabla_ih_j\bs^i\bs^j+\frac1{12}\bar{R}_{ij}\bs^i\bs^j],\\
    A_1&=\frac{-i}{32\pi^2}\frac{\bar{\De}^{1/2}(\spx,\spy)\frac16\bnabla^jf_{ji}\bs^i}{|\xi^2|^{1/4}(\spx)|\xi^2|^{1/4}(\spy)}\approx\frac{-i}{32\pi^2|\xi^2|^{1/2}(p)}\frac16\bnabla^jf_{ji}\bs^i,\\
    A_2&=\frac{i}{32\pi^2}\frac{\bar{\De}^{1/2}(\spx,\spy)}{|\xi^2|^{1/4}(\spx)|\xi^2|^{1/4}(\spy)}\frac1{12}\Sp(f^2+2E^{\br0})\approx\frac{i}{32\pi^2|\xi^2|^{1/2}(p)}\frac1{12}\Sp(f^2+2E^{\br0}),
\end{split}
\end{equation}
expanded near the midpoint (see Appendix \ref{Some_Expan}), we arrive at
\begin{equation}
    D^{\br+}(x,y)\approx\frac{-i}{32\pi^2}\Big\{[1+\frac{1}{12}R_{\mu\nu}(t\xi^\mu+\bs^\mu)(t\xi^\nu+\bs^\nu)]\frac{4M}{z_0}K_1(2Mz_0)-[\frac{\xi^2}{6}f^2+h^2-\frac23\nabla_\la h^\la]K_0(2Mz_0)\Big\}.
\end{equation}
Finally, we expand the Macdonald functions at the small argument \cite{GrRy},
\begin{equation}
    K_0(x)\approx-\ln\frac{x}2-\ga,\qquad K_1(x)\approx\frac1{x}+\frac{x}2(\ln\frac{x}2+\ga-\frac12),
\end{equation}
and obtain
\begin{equation}\label{pos_freq_func_s_main}
    D^{\br+}(x,y)\approx\frac{-i}{8\pi^2}\Big\{\frac{1+\frac{1}{12}R_{\mu\nu}\s^\mu\s^\nu}{\s}+\frac12(m^2-\frac16 R)\ln\frac{e^{2\ga} M^2\s}2-\frac{M^2}{2}\Big\},
\end{equation}
where $\ga$ is the Euler constant and, as before, $\s$ is the world function, $\s_\mu$ is its derivative.

We see that the singularities of the retarded and advance Green functions lie on the geodetic light cone as it should be according to the general theorems. The expansion of these functions in terms of $\s$ coincides with the standard one (see \cite{DeWGAQFT,BarvVilk}) and is independent of $\xi^\mu$. Whereas the imaginary part of the propagator depends on the Killing vector nontrivially and ``remembers'' the vacuum state with respect to which the propagator is defined. We also see that the correction to the mass squared is reproduced again. Without it, in particular, the factor at the logarithm in \eqref{pos_freq_func_s_main} would not have the the covariant form independent of $\xi^\mu$. The mention should be made that the positive-frequency function \eqref{pos_freq_func_s_main} cannot be employed to analyze the Unruh effect or the radiation reaction problem for particles (detectors) in a free fall. For these problems the main contribution to the integrals determining the force or transition rate (see, e.g., \eqref{trans_rate}) comes from a sufficiently large point separation of the order of the inverse acceleration $a^{-1}$ (do not confuse with $a(s)$ in Eq.\eqref{a(s)}). This is the so-called the radiation formation length (see \cite{Nikish}). In a free fall, $a$ is of the order of $L^{-1}$ and so the expansion bringing us to \eqref{pos_freq_func_s_main} under the assumption that $l\ll L$ are not valid. However, these expansions may be justified in the case of sufficiently large accelerations $a\gg L^{-1}$ caused by the external (nongravitational) forces. Then the terms entering the Unruh acceleration \eqref{effinterval} in the square brackets can be neglected.

Concluding this section, we write out the $s$-representation for the one-loop correction to the effective action induced by one bosonic mode. The general formula reads as (see, e.g., \cite{gmse,KalKaz})
\begin{equation}\label{one_loop_2}
    \Ga^{\br1}_{1b}/T=\int_0^\La\frac{d\omega}{2}\Sp\theta(H(-\omega))=\int d\spx\sqrt{\bar{g}}\int_0^\La d\omega\int\frac{ds}{4\pi is}G(-\omega,s;\spx,\spx),
\end{equation}
where the integration contour over $s$ runs a little bit lower than the real axis and $\La$ characterizes the energy cutoff. On the diagonal, the heat kernel is symmetric with respect to the replacement $\omega\rightarrow-\omega$ and so we have almost the same integral as for the positive-frequency function taken on the diagonal (cf. Eq. \eqref{pos_freq_func}). The energy cutoff can be conveniently imposed by introducing $t_-=-i\La^{-1}$ similar to the positive-frequency function. Then making the same approximations as in deriving \eqref{pos_freq_func_s}, we can write
\begin{multline}
    \Ga^{\br1}_{1b}=\int dx\sqrt{\bar{g}}\int_0^\infty d\omega\omega^{d/2}\int\frac{ds}{(4\pi is)^{d/2+1}}\det\Bigl(\frac{\sin s\ka}{s\ka}\Bigr)^{-1/2} e^{i\omega(S_0(\bs=0)-t_-)-is\bar{m_0}^2/\omega}=\\
    =i^{d-1}\int\frac{dx\sqrt{|g|}}{4\sqrt{|\xi^2|}}\int\frac{ds}{(4\pi)^{d/2}}\det\Bigl(\frac{\sin s\ka}{s\ka}\Bigr)^{-1/2} \Bigl(\frac{\bar{m}_0^2}{a}\Bigr)^{(d+2)/4}H^{(1)}_{d/2+1}(2\bar{m}_0a^{1/2}),
\end{multline}
where $\bar{m}_0$ equals $\bar{m}$ at $\bs=0$ and $a(s)$ is given by the expression \eqref{a(s)} with $\bs=0$ and $t_-=-i\La^{-1}$.

\section{Discussion}

In the previous sections we obtained an approximate solution to a purely mathematical problem of finding the Green function for a scalar quantum field on a stationary background under the assumption that the external fields varies slowly from point to point (the Boulware vacuum was implied). We have shown that the two-point Green function of a scalar field on a curved background and the induced effective action (in particular, its imaginary part, which cannot be removed by counterterms) depend nontrivially on the Killing vector $\xi^\mu$, i.e., this Killing vector field enters explicitly into the expressions and these expressions cannot be rewritten in a local form in terms of the $4$-metric and its curvature alone. This dependence stems from the fact that the mode functions of the field operators on a stationary background are chosen so as to diagonalize the quadratic part of the energy operator
\begin{equation}\label{energy}
    E_{\br2}[\phi_S,\pi_S]=\int_\Si d\Si_\mu T^\mu_{\br2\nu}[\phi_S,\pi_S]\xi^\nu,
\end{equation}
which explicitly depends on the Killing vector field. The operators $\phi_S$ and $\pi_S$ are the canonically conjugate Schrodinger field operators written in terms of the creation and annihilation operators, $T^{\mu\nu}$ is the (regularized) operator of the energy-momentum tensor, and the index 2 in \eqref{energy} reminds us that only the quadratic part of the operator in terms of $\phi_S$ and $\pi_S$ is retained. If a family of the hypersurfaces $\Si$ are the Cauchy surfaces then the operator \eqref{energy} coincides with the quadratic part of the evolution generator (the Hamiltonian). In the interaction picture, the field operators are constructed in terms of the mode functions which are stationary with respect to the Killing vector and such that the operator \eqref{energy} remains diagonalized. The bare vacuum state is defined as the state corresponding to the least eigenvalue of the energy operator. The dressed vacuum and particles states are determined by the full energy operator and not by its quadratic part, although in this paper we considered the free scalar field when \eqref{energy} coincides with the full energy. Note that for a flat spacetime with the vacuum state corresponding to the zero eigenvalue of the Hamiltonian the dependence on the Killing vector disappears, at least formally, because of the global Poincar\'{e} symmetry of the Schrodinger equation.

We have just briefly described the standard construction of quantum field theory on stationary backgrounds (see, e.g., \cite{DeWGAQFT,DeWQFTcspt}). This construction apart from the standard dynamical content (the Schrodinger or Heisenberg equations following from the form of the Lagrangian of the theory) involves one additional postulate, which tells us how to construct the Schrodinger operators in terms of the creation-annihilation operators and define thereby all the operators arising in the theory. Due to the ordering problems, different representations of the Schrodinger field operators may lead to inequivalent quantum field theories (see, e.g., \cite{Fulling}), though their classical dynamics coincide. Therefore, for non-stationary backgrounds, we also need the additional construction which singles out a unique (up to a well-defined unitary transform) representation of the Schrodinger operators at any instant of time. It is natural to suppose (see, e.g., \cite{GriMaMos}, Chap. 8) by analogy with the stationary case that the energy operator \eqref{energy} defined with respect to some vector field $\xi^\mu$ is diagonal at any instant of time. For stationary backgrounds this vector field should coincide with or be close to the Killing vector field. Notice that, in general, the so defined energy operator differs from the Hamiltonian for non-stationary backgrounds. Thus we shall have the Schrodinger field operators related by a unitary time-dependent transform at different instants of time (for mathematical details see, e.g., \cite{DerezBrun}). In order that the dynamical content of the theory remains intact, at least formally, we should transform the Hamiltonian accordingly to these nonstationary unitary transformations. Then the formal Heisenberg equations looks like the classical equations of motion. The ordering problems mentioned above are accompanied by the infrared and ultraviolet infinities, as for any local quantum field theory, and should be handled by the appropriate regularization and renormalization procedures.

In the case of a nonstationary metric, the energy momentum tensor operator is not covariantly divergenceless \cite{Hawk1,Grib}, in general. However, we can choose the vector field $\xi^\mu$ so as to remedy this defect demanding the fulfillment of the Ward identity. This results in the dynamical equations of a hydrodynamic form \cite{gmse} for the vector field $\xi^\mu$ with the appropriate boundary and initial conditions. The form of these equations strongly suggests that the field $g_\mu=\xi_\mu/\xi^2$ should be quantized. Then the whole construction is explicitly generally covariant. The quantization of the hydrodynamic equations is more or less known \cite{LandLifstat} and is used in describing the cosmological fluctuations (see, e.g., \cite{GorbRub,Mukhan}). The quantum field $\hat{g}_\mu$ is decomposed into the condensate part $g_\mu$ and the fluctuations $\de\hat{g}_\mu$. Then the latter are represented in terms of the creation and annihilation operators, the particles being identified with the phonons. The properties of these phonons follow from the fluid energy-momentum tensor. In particular, as found in \cite{gmse}, the sound speed squared of such phonons equals $-\vf_N/\al$ in the weak field limit, where $\vf_N$ is the Newtonian potential and $\al$ is the power of decay of the induced energy-momentum tensor at large distances from a gravitating body, i.e., $T_{\mu\nu}\sim r^{-\al}$. In terms of the Feynman diagrams, the phonons $\de\hat{g}_\mu$ appear in the graphs. To say figuratively, it looks like the particles ring in moving in the condensate of the field $\hat{g}_\mu$.

Although we do not touch in this paper the problem of quantization of the gravitational field, it should be noted that one or another solution to the time problem is necessary for the gravity quantization programme (for a review, see \cite{Isham}). The vector field $\xi^\mu$ and its dual, when integrable, can be employed to define the so-called world time \cite{LandLifshCTF} and provide the possible preferred definition of time. Such a method to solve this problem is not new and is known as the reference fluid approach (see, e.g., \cite{DeWQG,Isham,Kuchar,CiMoZo}). A distinctive feature of the approach proposed in \cite{gmse} is that we do not introduce such a fluid by hand, but deduce its equations of motion from the requirement of the covariant divergenceless of the energy-momentum tensor (for other similar approaches see, for example, \cite{Kostel,JacMat,Horav,Odints}). One of the purposes of the present paper was to show that the vector field defining such a fluid is already contained in the formalism of quantum field theory on a curved background. The existence of this vector (dynamical or not) allows one to define formally the gravitational and inertial forces as the $4$-vectors (see \cite{Zelman,LandLifshCTF} and Eqs. \eqref{geodesics}, \eqref{f_iner} below). Notice that for certain Euclidian approaches (see, e.g., \cite{BarVilkcov,WiMuZe,Shapir}), where the Killing vector does not appear, the induced effective action contains the terms of the form $R\nabla^{-2}R$ which are essentially nonlocal. These terms involve the inverse powers of the covariant d'Alambertian and the correct definition of such constructions requires the assignment of the initial and boundary conditions. In fact, this is equivalent to the introduction of the new fields into the model which also contribute to the covariant divergence of the energy-momentum tensor.

Observe also a remarkable feature of the construction described above -- it is invariant with respect to uniform dilatations of the Killing vector $\xi^\mu$ (see, e.g., \cite{DeWGAQFT,LandLifshCTF}). This demonstrates the invariance of a theory with respect to a possible choice of the energy unit. In order to maintain the correspondence between the energy and the Hamiltonian for stationary backgrounds, the time variable parameterizing the Cauchy surfaces should be also stretched in such a way that the product ``energy''$\times$``time'' will be invariant. Therefore we shall call this symmetry as the energy-time symmetry. One may check that all the observables appearing in this paper are invariant under such a transform. However, if the theory possesses an intrinsic scale, say, the cutoff scale, this invariance is violated. Formally, the violation follows from the fact that the right-hand side of \eqref{energy} and other composite operators must be regularized. As a result, the additional terms involving $g_\mu$ appear and break the symmetry. The typical terms violating the energy-time symmetry are presented in Eq. (21) of \cite{gmse} (see also \cite{DeWQFTcspt}). One can distinguish two types of such violating terms: hard violating and soft violating. The former scale as certain powers under dilations of the Killing vector, while the latter do it logarithmically. The coefficients at the soft violating terms are well known and nothing but the scaling functions ($\be$ or $\ga$) of the renormalization group taken at that scale where the model is in a perturbative regime (if it exists). Note that all the particles of the model, even with large masses, contribute to this coefficients on equal footing (for a concrete calculations in the standard model see \cite{KarKaz}). The relation of these coefficients with the scaling functions is not surprising (see, e.g., \cite{BrOtPa,FrZel}) as long as the energy-time symmetry involves a dilatation. Hence, if the theory were conformal on a quantum level, such anomalous terms would not arise. The restriction on the parameters of the models implying $\be=0$ can be found, for example, in \cite{tHooft}. According to \cite{gmse}, the anomaly of the effective action under the energy-time dilatations defines the enthalpy density of the condensate $g_\mu$, although the hydrodynamic equations for $g_\mu$ are not empty even in the case of a vanishing enthalpy.

%\newpage
\appendix
\section{$(3+1)$ with the Killing vector}\label{4_Kill}

In this appendix, we collect some formulas regarding the differential calculus on the Riemannian manifold with the Killing vector. If the metric possesses the Killing vector $\xi^\mu$ then the following useful relations hold
\begin{equation}
\begin{gathered}
    f_{\mu\la}g^\la=0,\qquad g_\la h^\la=0,\qquad
    g^\la\nabla_\la f_{\s\mu}=g_{[\mu}f_{\s]\la}h^\la,\\
    g^{\mu}g^{\nu}\nabla_{\mu}h_{\nu}=g^2h^2,
    \qquad\nabla^\la f_{\la\mu}g^\mu=-\frac12f^2,\qquad\nabla_\mu g_\nu=\frac12 f_{\mu\nu}-h_{(\mu}g_{\nu)},
\end{gathered}
\end{equation}
and for the curvature
\begin{equation}
\begin{gathered}
    g^\la R_{\la\nu\s\mu}=\frac12\nabla_\nu f_{\s\mu}-\frac12h_{[\s}f_{\mu]\nu}+h_{[\s}g_{\mu]}h_\nu+f_{\s\mu}h_\nu-g_{[\s}\nabla_{\mu]}h_\nu,\\
    R_{\mu\nu}g^\mu g^\nu=\frac14 f^2-g^2\nabla_\la h^\la,\qquad
    g^\la R_{\la\mu}=f_{\mu\la}h^\la-\frac12\nabla^\la f_{\la\mu}-g_\mu\nabla_\la h^\la,
\end{gathered}
\end{equation}
where $f_{\mu\nu}=\partial_{[\mu}g_{\nu]}$, $h_\mu=\partial_\mu\ln\sqrt{|\xi^2|}$, and $f^2:=f_{\mu\nu}f^{\mu\nu}$. The latter notation is used for scalars. As for matrices, $f^2:=f_{\mu\la}f^\la_{\ \nu}$ and then $\Sp f^2=-f^2$.

Using the Killing vector we can construct the projected connection,
\begin{equation}
    \bar{\Ga}^\al_{\mu\be}:=\Ga^\al_{\mu\be}+\frac{\xi^2}{2}(f^\al_{\ (\mu}g_{\be)}+2h^\al g_\mu g_\be-2g^\al h_{(\mu}g_{\be)}),
\end{equation}
such that
\begin{equation}
\begin{gathered}
    \bnabla_\mu \bar{g}_{\rho\s}=0,\qquad\bnabla_\mu\xi^\nu=0,\qquad\bnabla_\mu g_\nu=\frac12 f_{\mu\nu},\\
    \bnabla_\mu g_{\rho\s}=\frac{\xi^2}{2}f_{\mu(\rho}g_{\s)}+2\xi^2h_\mu g_\rho g_\s,\qquad\bnabla_\mu \bar{g}^{\rho\s}=-\frac{\xi^2}{2}f_\mu^{\ (\rho}g^{\s)},\qquad
    \bnabla_\rho\bar{g}^{\rho\s}=0.
\end{gathered}
\end{equation}
This connection can be also obtained from the Levi-Civita connection for the metric
\begin{equation}
    G_{\mu\nu}:=g_{\mu\nu}-\xi^2g_\mu g_\nu+\la^{-1}g_\mu g_\nu,
\end{equation}
in the limit $\la\rightarrow\infty$ (see \cite{KalKaz}). Then the equation for geodesics of the metric $g_{\mu\nu}$ takes the form (cf. \cite{LandLifshCTF}, Sec. 88)
\begin{multline}\label{geodesics}
    \nabla_uu^\mu=\bnabla_uu^\mu-(\xi u)(f^\mu_{\ \nu}u^\nu+(gu)h^\mu-2(hu)g^\mu)=0\;\Rightarrow\\
    \bar{g}_{\mu\nu}\nabla_uu^\nu=\bnabla_u(\bar{g}_{\mu\nu}u^\nu)-(\xi u)(f_{\mu \nu}u^\nu+(gu)h_\mu)=0,
\end{multline}
The second term on the left-hand side of the first equation can be regarded as the acceleration caused by a stationary gravitational field \cite{Zelman}. The first term in parentheses is associated with the so-called gravimagnetic force, whereas the second term is responsible for the gravielectric component of the force. The factor at these parentheses is the energy of the particle divided by its mass. In particular, the vector field $\xi^\mu$ allows one to define formally the inertial forces as the $4$-vector
\begin{equation}\label{f_iner}
    f_{iner}^\mu:=-m[\nabla_uu^\mu+(\xi u)(f^\mu_{\ \nu}u^\nu+(gu)h^\mu)-2(hu)g^\mu]\;\Leftrightarrow\;f_{iner}^\mu+f_g^\mu+f^\mu=0,
\end{equation}
where $f^\mu:=m\nabla_uu^\mu$ and $f_g^\mu:=m(\xi u)(f^\mu_{\ \nu}u^\nu+(gu)h^\mu-2(hu)g^\mu)$ are the external (nongravitational) and the gravitational forces, respectively.

The Riemann and Ricci curvatures of the projected connection read as (see also \cite{KalKaz})
\begin{equation}
\begin{split}
    \bar{R}^\rho_{\ \nu\s\mu}&=R^\rho_{\ \nu\s\mu}-\frac{\xi^2}{2}\Bigl(f_{\s\mu}f_\nu^{\ \rho}-g_{[\s}\nabla_{\mu]}f_{\nu}^{\ \rho}-g_\nu\nabla^\rho f_{\s\mu}-\frac12 f_{\nu[\s}f_{\mu]}^{\ \rho} \Bigr)-\\
    &+\xi^2\Bigl(g_{[\s}h_{\mu]}h_\nu g^\rho-g_{[\s}h_{\mu]}h^\rho g_\nu+\frac12h_{[\s}f_{\mu]\nu}g^\rho-\frac12g_{[\s}f_{\mu]\nu}h^\rho+g_{[\s}\nabla_{\mu]}h_\nu g^\rho-g_{[\s}\nabla_{\mu]}h^\rho g_\nu-\\
    &-f_{\s\mu}(h_\nu g^\rho-h^\rho g_\nu)
    -f_\nu^{\ \rho} h_{[\s}g_{\mu]}-\frac12 h_{[\s}f_{\mu]}^{\ \rho}g_\nu+\frac12 g_{[\s}f_{\mu]}^{\ \rho}h_\nu \Bigr)+\frac{\xi^4}4 g_\nu g_{[\s}f_{\mu]}^{\ \la}(f_{\la}^{\ \rho}+2h_\la g^\rho),\\
    \bar{R}_{\mu\nu}&=R_{\mu\nu}+h_{\mu}h_{\nu}+\nabla_{\mu}h_{\nu}-\frac{\xi^2}{2}[f_{\mu\rho}f^{\rho\ }_{\ \nu}+g_{(\mu}\nabla_{\rho}f_{\nu)\ }^{\ \rho}+2g_{\mu}g_{\nu}(h^2-\nabla_{\rho}h^{\rho})+g_{(\mu}f_{\nu)}^{\ \rho}h_{\rho}]+\frac{\xi^4}4g_{\mu}g_{\nu}f^2.
\end{split}
\end{equation}
The scalar curvature becomes
\begin{equation}
    \bar{R}=\bar{g}^{\mu\nu}\bar{R}_{\mu\nu}=R+2\nabla_{\rho}h^{\rho}+\frac{\xi^2}{4}f^2.
\end{equation}
Besides, we have
\begin{equation}\label{hh}
    \bnabla^ih_i=\bar{g}^{\mu\nu}\bnabla_\mu h_\nu=\nabla_\la h^\la-h^2,\qquad h_ih^i=h^2,\qquad f_{ij}f^{ij}=f^2.
\end{equation}
Also we need the quantities in the adapted coordinates in the weak field limit \cite{LandLifshCTF}, Sec. 105,
\begin{equation}
\begin{gathered}
    \bar{R}_{iklm}\approx\frac{r_g}{2r^3}(2\de_{i[l}\de_{m]k}-3n_in_{[l}\de_{m]k}+3n_kn_{[l}\de_{m]i}),\qquad \bar{R}_{km}\approx\frac{r_g}{2r^3}(\de_{km}-3n_kn_m),\\
    \bar{R}_{ik}\bar{R}^k_j\approx\frac{r_g^2}{4r^6}(\de_{ij}+3n_in_j),\qquad \bar{R}_{iklm}\bar{R}_j^{\ klm}\approx\frac{r_g^2}{2r^6}(5\de_{ij}-3n_in_j),\\
    \bnabla^2\bar{R}_{ij}dx^idx^j=\frac{3r_g^2}{2r^6}(1-\frac{r_g}{r})^{-1}dr^2-\frac{3r_g^2}{4r^6}(r^2d\theta^2+r^2\sin^2\theta d\vf^2)\approx-\frac{3r_g^2}{4r^6}(\de_{ij}-3n_in_j)dx^idx^j,
\end{gathered}
\end{equation}
where $n_i=x_i/r$ and $\bnabla^2 \bar{R}_{ij}$ is calculated for the Schwarzschild metric (see \cite{LandLifshCTF}, Sec. 100). Recall that $\bar{R}=0$ and $\nabla^\la h_\la=0$ for this metric.

\section{Some expansions}\label{Some_Expan}

In this appendix, we give the expansions of some quantities appearing in the main text. For the metric in the Riemann normal coordinates we have \cite{Petrov,BekPark}
\begin{equation}
\begin{gathered}
    \bar{g}_{ij}=\de_{ij}-\frac13\bar{R}_{ikjl}y^ky^l-\frac1{3!}\bnabla_s\bar{R}_{ikjl}y^ky^ly^s+\frac1{5!}(\frac{16}3\bar{R}_{sjn}^{\ \ \ m}\bar{R}_{likm}-6\bnabla_{sn}\bar{R}_{ikjl})y^ky^ly^sy^n+O(y^5),\\
    -\bar{g}=1-\frac13\bar{R}_{ij}y^iy^j-\frac{1}{3!}\bnabla_i \bar{R}_{jk}y^iy^jy^k-\frac1{4!}(\frac65\bnabla_{ij}\bar{R}_{kl}+\frac4{15}\bar{R}_{mijn}\bar{R}^{m\ \ n}_{\ \ kl}-\frac43\bar{R}_{ij}\bar{R}_{kl})y^iy^jy^ky^l+O(y^5).
\end{gathered}
\end{equation}
Also we deduce
\begin{equation}
\begin{gathered}
    |\xi^2|^{-1/4}(\spx)|\xi^2|^{-1/4}(\spy)=|\xi^2|^{-1/2}(p)(1-\frac18\bnabla_i h_j\bs^i\bs^j+O(\bs^3)),\\
    \bar{\De}^{-1}(\spx,\spy)=1-\frac16\bar{R}_{ij}\bs^i\bs^j+O(\bs^3).
\end{gathered}
\end{equation}
In order to prove the relation \eqref{quasiint_exp}, the following expansions may be useful
\begin{equation}
\begin{split}
    -1&\equiv\xi^2(u^0)^2+u_iu^i+2\big[\xi^2((u^0)^2h_iu^i+u^0\dot{u}^0)+u_i\dot{u}^i\big]\tau+\\
    &+\Big\{\xi^2\big[(u^0)^2(h_i\dot{u}^i+2(h_iu^i)^2+\bnabla_ih_ju^iu^j)+4u^0\dot{u}^0h_iu^i
    +(\dot{u}^0)^2+u^0\ddot{u}^0+\\
    &+\frac{u^0}2f_{ij}u^i\dot{u}^j \big]+(\dot{u}^i)^2+u_i\ddot{u}^i\Big\}\tau^2+O(\tau^3),\\
    (\nabla_uu)^2&=\xi^2(\dot{u}^0)^2+(\dot{u}^i)^2+\xi^2\Big[\xi^2(u^0)^2\big(h^2(u^0)^2+2u^0h^if_{ij}u^j-u^if^2_{ij}u^j\big)-\\
    &-2u^0(\dot{u}^if_{ij}u^j+u^0h_i\dot{u}^i)+4h_iu^iu^0\dot{u}^0+4(h_iu^i)^2(u^0)^2 \Big]+O(\tau),\\
    \xi^2(p)&=\xi^2\Big[1+h_iu^i\tau+\big(h_i\dot{u}^i+(h_iu^i)^2+\frac12\bnabla_ih_ju^iu^j\big)\frac{\tau^2}2\Big]+O(\tau^3).
\end{split}
\end{equation}
Here all the fields are taken at the origin of the system of coordinates described after Eq. \eqref{natural_param}. Notice that $u^0=g_\mu u^\mu$ at the origin of this frame.

\section{Conformal transformations}\label{Conf_Trans}

In this appendix, we discuss an influence of the conformal transformations on the form of the approximate expression for the heat kernel \eqref{HK_resum} and, hence, on the form of the positive-frequency function \eqref{pos_freq_func_1}. For a static space-time, the method of conformal transformations is a standard tool to obtain the approximate expression for the propagator (see \cite{Page}) of a conformal scalar field. Using this transform, one passes from the static to the ultrastatic metric (the optic metric) so that the time variable decouples. Then the standard technique \cite{Schwing,Parker,Avramid,VasilHeatKer,BekPark} is applied to the operator of Laplacian type corresponding to the Euclidean sector of the optic metric and the approximate expression for the heat kernel is constructed thereby. After that the inverse conformal transformation is performed what leads to a certain approximate expression for the heat kernel associated with the initial operator.

Consider the conformal (Weyl) transform of the stationary metric
\begin{equation}
    g_{\mu\nu}=e^{2\Omega}\tilde{g}_{\mu\nu},
\end{equation}
where $\Omega$ is independent of time. The Killing vector is not transformed. Direct calculations show that formula \eqref{Hamiltonian} changes accordingly ($d=3$):
\begin{equation}\label{Hamiltonian_conf}
\begin{split}
    H(x,y)&=:|\tilde{\xi}^2|^{-1/4}(\spx)\int\frac{d\omega}{2\pi}e^{-i\omega t}\tilde{H}(\omega;\spx,\spy)|\tilde{\xi}^2|^{-1/4}(\spy)e^{-2\Omega(\spy)},\\
    \tilde{H}(\omega;\spx,\spy)&=\biggl\{(\tilde{\bar{\nabla}}_i+i\omega g_i)\tilde{\bar{g}}^{ij}(\tilde{\bar{\nabla}}_i+i\omega g_i)-e^{2\Omega}[\frac{\omega^2}{\xi^2}+m^2+\kappa R+\frac12\nabla^\mu (h_\mu+\nabla_\mu\Omega)-\\
    &-\frac14g^{\mu\nu}(h_\mu+\nabla_\mu\Omega)(h_\nu+\nabla_\nu\Omega)]\biggr\}\frac{\de(\spx-\spy)}{\tilde{\bar{g}}^{1/4}(\spx)\tilde{\bar{g}}^{1/4}(\spy)},
\end{split}
\end{equation}
where we have introduced the nonminimal coupling term $\kappa R$ and used formulas \eqref{hh}. Note that the Tolman temperature one-form $g_i$ is invariant under the conformal transformations. We see that the different choices of the function $\Omega$ result in the different perturbation theories for the heat kernel. The main goal of these conformal transformations is to rearrange the perturbation theory in such a way that the higher vertices give a negligible contribution to the heat kernel. The vertices are determined by the potential in \eqref{Hamiltonian_conf} with the correction \eqref{grav_pot} and by the terms coming from the expansion of the kinetic term (the first term in \eqref{Hamiltonian2}) in a covariant Taylor series. As for \eqref{grav_pot} is concerned, we write out here only the leading contribution
\begin{equation}\label{quan_corr}
    -\frac16\tilde{\bar{R}}=-\frac{e^{2\Omega}}{6}[R+2\nabla^\mu(2\nabla_\mu\Omega+h_\mu)-2(\nabla_\mu\Omega+2h_\mu)\nabla^\mu\Omega+\frac{\xi^2}{4}f^2].
\end{equation}
For the optic metric $\nabla_\mu\Omega=h_\mu$ and so all the terms with $h_\mu$ entering $E_0$ (see Eq. \eqref{Ham_func}) are canceled out. At $\kappa=1/6$, the scalar curvature also disappears from $E_0$ and the mass squared acquires the correction $-\xi^2 f^2/24$.

Let us roughly estimate the corrections to the heat kernel due to vertices in the Hamiltonian \eqref{Hamiltonian_conf} represented in the form \eqref{Hamiltonian2}. For the optic metric the correlators appearing in the perturbation theory are of the order
\begin{equation}
    \lan xx\ran\sim\{(m^2\partial^2\xi^2)^{-1/2},(\partial^2R)^{-1/2}\},\qquad\lan xp\ran\sim\{1,1\},\qquad \lan pp\ran\sim\{(m^2\partial^2\xi^2)^{1/2},(\partial^2R)^{1/2}\},
\end{equation}
where the left case is for massive particles and the right case is for massless ones. Then, a comparison of the contribution from the vertex with the quadratic part gives
\begin{equation}
    \frac{m^2\partial^4\xi^2\lan xx\ran\lan xx\ran}{m^2\partial^2\xi^2\lan xx\ran}\sim\frac1{\xi mL},\qquad \frac{R\lan xx\ran\lan pp\ran}{\partial^2R\lan xx\ran}\sim\frac{R}{(\partial^2R)^{1/2}}\sim \e^{1/2},
\end{equation}
where we have used the notation from Eqs. \eqref{scales_weak}, \eqref{scales_strong}. We see that in the massive case the perturbation theory is reliable almost everywhere $\xi mL\gg1$, while for massless particles it is valid only for a sufficiently weak field $\e\ll1$. To be certain that the correction to the mass squared ($-\xi^2 f^2/24$) does exist, we need to consider the massless case, where this correction could be relevant, and check that this contribution is not overridden by the loop corrections. Besides, the model must be in a perturbative regime. Only then can we say that such a correction is reliable. In the massless case and in the weak field limit we should compare
\begin{equation}
    \xi^2 f^2\sim\frac{\e^2}{L^2}\quad\text{and}\quad\partial^2R\lan xx\ran\sim(\partial^2R)^{1/2}\sim\frac{\e^{1/2}}{L^2}.
\end{equation}
Thus we see that this correction to the mass squared is not reliable.

For the metric $\bar{g}_{ij}$ ($\Omega=1$) the correlators can be estimated as follows
\begin{equation}
    \lan xx\ran\sim (\omega^2\partial^2\xi^{-2})^{-1/2},\qquad\lan xp\ran\sim 1,\qquad \lan pp\ran\sim (\omega^2\partial^2\xi^2)^{1/2}.
\end{equation}
Hence, the estimates for the relative contributions of vertices become
\begin{equation}\label{applicab1}
    \frac{\omega^2\partial^4\xi^{-2}\lan xx\ran\lan xx\ran}{\omega^2\partial^2\xi^{-2}\lan xx\ran}\sim\frac{c}{\xi^2\omega L},\qquad\frac{\bar{R}\lan xx\ran\lan pp\ran}{\omega^2\partial^2\xi^{-2}\lan xx\ran}\sim\frac{\e^{1/2}}{\omega L},
\end{equation}
where we take into account that $\bar{R}\sim\xi^{-2}L^{-2}$ (\cite{LandLifshCTF}, Sec. 100) and $\bar{R}\sim\e L^{-2}$ in the strong and weak field limits, respectively. The first estimation is written for the vertices coming from the expansion of $\omega^2/\xi^2$ in the potential and the constant $c$ is a numerical coefficient at the corresponding vertex. The second estimation is given for the vertices standing at momenta: $\partial^{n-2}\bar{R}x^npp$. To verify that the correction to the mass squared $-h^2/4$ is reliable we should compare
\begin{equation}
    h^2\sim\{\xi^{-4}L^{-2},\e^2L^{-2}\}\quad\text{and}\quad\omega^2\partial^2\xi^{-2}\lan xx\ran\sim\{\frac{\omega}{\xi^2L},\frac{\omega\e^{1/2}}{L}\},
\end{equation}
where the left case is for the strong field limit ($\xi^2\rightarrow0$) and the right case is for the weak field limit. Demanding that $h^2$ must be larger than or comparable with the quantum correction and combining this condition with the requirements of applicability of the loop expansion (the right-hand sides of \eqref{applicab1} must be small), one deduces
\begin{equation}\label{applicab_weak}
    \omega L\lesssim\e^{3/2},\qquad c\ll\e^{1/2}\omega L\sim\e^2,\qquad\e^{1/2}\ll\omega L\sim\e^{3/2},
\end{equation}
for the weak field limit, and
\begin{equation}
    \xi^2\omega L\lesssim1,\qquad c\ll \xi^2\omega L,\qquad\omega L\gg 1,
\end{equation}
for the strong field one. Thus we see that the mass correction is not reliable in the weak field limit, but for the strong fields and small $c$ there is a regime where this correction dominates and so it is reliable. The smallness of $c$ can be guaranteed in the case when the function $\omega^2/\xi^2$ entering the potential is accurately approximated by the first three terms of the covariant Taylor expansion (see Eq. \eqref{Eij}).

As for the weak field limit, we see that sufficiently large contributions come from the vertices standing at momenta $\partial^{n-2}\bar{R}x^npp$ and it is their contributions which violate \eqref{applicab_weak} (the last inequality). Fortunately, we can remove such vertices applying the conformal transformation to the Schwarzschild metric. It is well-known (\cite{LandLifshCTF}, Sec. 100) that the spatial part of the Schwarzschild metric $\bar{g}_{ij}$ is conformally flat
\begin{equation}
    \bar{g}_{ij}dx^idx^j=(1-\frac{r_g}{r})^{-1}dr^2+r^2d\Omega^2=(1+\frac{r_g}{4\rho})^4(d\rho^2+\rho^2d\Omega^2),\qquad r=\rho(1+\frac{r_g}{4\rho})^2.
\end{equation}
Therefore, performing the conformal transformation of the spacetime metric $g_{\mu\nu}$ with the conformal factor leading to the flat metric in the spatial sector, we remove all the vertices with the curvatures $\tilde{\bar{R}}$. The conformal factor can be conveniently written as
\begin{equation}
    \Omega=-2\ln\frac{1+|\xi^2|^{1/2}}{2},\qquad\nabla_\mu\Omega=-2\frac{|\xi^2|^{1/2}h_\mu}{1+|\xi^2|^{1/2}},
\end{equation}
where we assume that the Killing vector is normalized to unity at spatial infinity. Substituting this expressions to \eqref{Hamiltonian_conf} and \eqref{quan_corr}, we find
\begin{equation}
    \frac12\nabla^\mu(h_\mu+\Omega_\mu)-\frac14(h_\mu+\Omega_\mu)^2=\frac12\frac{1-|\xi^2|^{1/2}}{1+|\xi^2|^{1/2}}\nabla^\mu h_\mu-\frac14h^2,\qquad\tilde{\bar{R}}=e^{2\Omega}\Big(R+2\frac{1-3|\xi^2|^{1/2}}{1+|\xi^2|^{1/2}}\nabla^\mu h_\mu\Big),
\end{equation}
where we put $f^2=0$. Now we take into account that $R=0$ and $\nabla^\mu h_\mu=0$ for the Schwarzschild metric. Consequently, the mass squared acquires the correction $-h^2/4$ again. We shall not elaborate here a further analysis of the heat kernel with this metric, but note that it is very plausible that the correction $-h^2/4$ exists even in the weak field limit.

%\newpage
\begin{acknowledgments}

The work is supported by the RFBR grant 12-02-31071-mol-a and by the Russian Ministry of Education and Science, contracts No. 14.B37.21.0911 and No. 14.B37.21.1298. I appreciate the anonymous referee of \cite{gmse} who pointed out the possible quantum character of the vector field $\xi^\mu$. I am also grateful to V.~G. Bagrov for discussions on the subject, and A.~A. Sharapov and S.~L. Lyakhovich for the constructive criticism. Of course, these people are in no way responsible for the possible flaws in the present paper.

\end{acknowledgments}

\end{document}